\documentclass[review,square,comma,sort&compress]{elsarticle}
\usepackage{graphicx}
\usepackage[figurename=Fig.,bf]{caption}
\usepackage{epsfig}
\usepackage{subfig}
\usepackage{amsmath}

\begin{document}
\begin{frontmatter}

\address[famu]{Florida A\&M University, Department of Physics,
Tallahassee FL,32307}

\title{Quantum toroidal moments of an elliptic toroidal helix in a constant magnetic field}

\author [famu]{J. {Williamson}\corref{cor1}}
\ead{johnny1.williamson@famu.edu}
\author [famu]{M. Encinosa}
\ead{mario.encinosa@famu.edu}

\cortext[cor1]{Corresponding Author}

\begin{abstract}
An effective one-dimensional Schr\"odinger equation for a spinless
 particle  constrained to motion near a toroidal helix immersed in an arbitrarily
 oriented constant magnetic field is developed.
The dependence of the induced toroidal moments on the magnetic flux through the helix is presented. The magnitude of the moments depend strongly on the component of the field normal to the toroidal plane.
A strong dependence on coil eccentricity is also indicated. It is also shown that field-curvature coupling potential terms are necessary to preserve the Hermiticity of the minimal prescription Hamiltonian.

\end{abstract}

\begin{keyword}
elliptic toroidal helix \sep toroidal moment \sep curvature potential \sep magnetic curvature potential \sep metaparticle
\end{keyword}

\end{frontmatter}

\bibliographystyle{elsarticle-num}

\section{Introduction}
Metamaterials comprise two and three dimensional grids of identical constituents sometimes referred to as metaparticles. To date,
much of the focus on metaparticle modeling has been on the macroscopic response of the bulk
array to electromagnetic radiation \cite{kaelscience,papas,marinov}. However, when metaparticles are eventually fabricated on scales at which quantum effects become non-negligible, quantum methods will have to be employed to capture the physics of the metaparticles, their mutual interactions, and their interactions with external fields.

As emphasized in \cite{marinov,williamson,papas}, there are both fundamental and practical reasons to investigate metaparticles that support  toroidal moments, of which a toroidal helix (TH) is one example\cite{afanasiev, ceulemans,
dubovik, papas, naumov,spaldin, sawada}. This paper serves as an extension of previous work \cite {williamson} wherein the three dimensional Schrodinger equation for a spinless particle in the region of a TH was reduced to an effective one dimensional equation and toroidal moments corresponding to system eigenstates were calculated. Here, a constant magnetic field (on the order of 1 Tesla) with arbitrary orientation is applied to the helix to study the induced toroidal moments as a function of field strength, orientation, and coil eccentricity.

The introduction of the field requires accounting for its effect in the one dimensional reduced model. It is not enough to simply evaluate the value of the vector potential for a given field on the points of the wire. Self-consistency with
the well known procedure by which curvature potentials arise from dimensional reduction \cite{chapblick, dacosta1,dacosta2,duclosexner,ee1,ee2,jenskoppe,matsutani1,matsutani2,taira,schujaff,burgsjens}
 requires a careful accounting of the coil curvature-field coupling.  A prescription exists for the three to two dimensional reduction which is extended here with suitable modification to arrive at the effective one dimensional magnetic interaction \cite{encarbB}.
  It is important to allow for arbitrary field orientation because of the lack of up-down
symmetry and the chiral nature of a TH; the inclusion of a magnetic field with a component tangent to the coil plane, along with a z-component complicates matters substantially.

The remainder of this paper is organized into five sections.
In the interest of keeping this paper relatively self-contained, section 2 presents
a parameterization for an $\omega$ turn TH. The Frenet system appropriate to the elliptic toroidal helix is introduced, and the three dimensional Hamiltonian $H^3_{\omega}$ is written. The vector potential in the Frenet system is determined. Section 3 is devoted to the reduction of $H^3_{\omega}$ to a one-dimensional Hamiltonian $H^1_{\omega}$. The one dimensional effective vector potential for use in $H^1_{\omega}$ is derived, and a dimensionless form of the one-dimensional Schr\"odinger equation is written. Section 4 motivates the basis set used in the Schr\"odinger equation, and presents the equations used to compute the Hamiltonian matrix elements and the toroidal moments. Section 5 reports results for toroidal moments as a function of coil eccentricity, field strength and field orientation. Finally, section 6 is reserved for conclusions and remarks concerning future work.

\section{The 3D ETH Hamiltonian subject to a constant B-field}
Consider an $\omega$-turn elliptic toroidal helix (ETH), of major radius $R$, minor horizontal radius $a$, and minor vertical radius $b$, parametrized by the Monge form
\begin{equation}
{\bf r}(\phi)=W(\phi)\hat{{\boldsymbol {{\boldsymbol {\rho}}}}}+b~{\rm sin}(\omega\phi)\hat{{\bf{k}}},
\end{equation}
where
\begin{equation}
W(\phi)=R+a~{\rm cos}(\omega\phi),
\end{equation}
and $\phi$ is the standard cylindrical coordinate azimuthal angle. The Hamiltonian for a particle allowed to move in the vicinity of the ETH in a magnetic field described by vector potential {\bf A}, using the minimal prescription, is
\begin{equation}
H^3_{\omega} = \frac{1}{2 {\textit m}_e} \big( \frac{\hbar} {\textit i}~\vec{\nabla} + e {\bf A} \big)^2,
\end{equation}
where $e$ is the magnitude of the electron charge, and $m_e$ is the electron mass.
In the Coulomb gauge, the Schr\"{o}dinger equation in three dimensions becomes
\begin{equation}
\bigg( \nabla ^2 + 2{\textit i} \frac{e}{\hbar}~{\bf A} \cdot \vec{\nabla} - \frac{e^2}{\hbar^2} {\bf A}^2 +  \frac{2 {\textit m}_e E}{\hbar^2}  \bigg) \Psi = 0.
\end{equation}
It is necessary to determine the gradient operator when considering the ambient space in which the ETH is embedded; therefore, an orthonormal coordinate system (Frenet system) is attached to the coil at every point along the ETH. The unit vector tangent to the ETH is
\begin{equation}
\hat{\bf T} = \frac{d{\bf r}(\phi)}{d\phi} ~{\bigg| \bigg| \frac{d{\bf r}(\phi)}{d\phi} \bigg| \bigg|}^{-1}.
\end{equation}
It is convenient to define 
\begin{equation}
f(\phi) \equiv {\bigg| \bigg| \frac{d{\bf r}(\phi)}{d\phi} \bigg| \bigg|}
\end{equation}
so that the Frenet-Serret relations can be written
\begin{equation}
\frac{d\hat{\bf T}}{d\phi} = f(\phi) \kappa(\phi) \hat{\bf N}
\end{equation}
\begin{equation}
\frac{d\hat{\bf N}}{d\phi} = -f(\phi) \kappa(\phi) \hat{\bf T} + f(\phi) \tau(\phi) \hat{\bf B}
\end{equation}
\begin{equation}
\frac{d\hat{\bf B}}{d\phi} = -f(\phi) \tau(\phi) \hat{\bf N}.
\end{equation}
The curvature and torsion of the space curve ${\bf r}(\phi)$ are represented by $\kappa(\phi)$ and $\tau(\phi)$ respectively.
Once the Frenet system has been established, two lengths, $q_{\mbox{\tiny N}}$ and $q_{\mbox{\tiny B}}$, are introduced along the normal (${\hat{\bf N}}$) and binormal (${\hat{\bf B}}$) directions, perpendicular to the ETH curve. The position vector for the particle near an ETH can now be written
\begin{equation}
{\bf x}(\phi,q_{\mbox{\tiny N}},q_{\mbox{\tiny B}})={\bf r}(\phi) + q_{\mbox{\tiny N}} \hat{\bf N} + q_{\mbox{\tiny B}} \hat{\bf B}.
\end{equation}
The requirement
\begin{equation}
d{\bf x} \cdot \vec{\nabla} = d{\phi}\frac{\partial}{{\partial \phi}} + d{q_{\mbox{\tiny N}}}\frac{\partial}{{\partial q_{\mbox{\tiny N}}}} + d{q_{\mbox{\tiny B}}}\frac{\partial}{{\partial q_{\mbox{\tiny B}}}},
\end{equation}
leads to the gradient operator
\begin{equation}
\begin{split}
\vec{\nabla} = \hat{{\bf T}} \frac{1}{f(\phi)(1-q_{\mbox{\tiny N}} \kappa(\phi))} \frac{\partial}{\partial \phi}+ \bigg( \hat{{\bf N}}~+~\hat{{\bf T}} \frac{q_{\mbox{\tiny B}} \tau(\phi)}{1-q_{\mbox{\tiny N}} \kappa(\phi)} \bigg) \frac{\partial}{\partial q_{\mbox{\tiny N}}} ~~\\ +~\bigg( \hat{{\bf B}}~-~\hat{{\bf T}} \frac{q_{\mbox{\tiny N}} \tau(\phi)}{1-q_{\mbox{\tiny N}} \kappa(\phi)} \bigg)\frac{\partial}{\partial q_{\mbox{\tiny B}}}~.
\end{split}
\end{equation}
The applied magnetic field in this work can be parametrized via
\begin{equation}
{\bf B} = B_\rho \hat{\boldsymbol \rho}_{\mbox{\tiny M}} + B_z \hat{\bf k},
\end{equation}
where $\hat{\boldsymbol \rho}_{\mbox{\tiny M}}$ indicates the direction of the magnetic field in the x-y plane at the angle $\phi = \phi_{\mbox{\tiny M}}$, i.e.
\begin{equation}
\hat{\boldsymbol \rho}_{\mbox{\tiny M}} = {\rm cos}(\phi_{\mbox{\tiny M}}) \hat{\bf i} +{\rm sin}(\phi_{\mbox{\tiny M}}) \hat{\bf j}.
\end{equation}Although its strength and direction may be chosen arbitrarily, the magnetic field remains constant once its properties are selected. The vector potential is calculated using
\begin{equation}
{\bf A}(\phi,q_{\mbox{\tiny N}},q_{\mbox{\tiny B}}) = \frac{1}{2}~{\bf B} \times {\bf x}(\phi,q_{\mbox{\tiny N}},q_{\mbox{\tiny B}}).
\end{equation}
To simplify the calculation of the ${\bf A} \cdot \vec{\nabla}$ term in Eq.(4), the vector potential can be expressed in the Frenet basis as
\begin{equation}
{\bf A}(\phi,q_{\mbox{\tiny N}},q_{\mbox{\tiny B}}) = A_{\mbox{\tiny T}}(\phi,q_{\mbox{\tiny N}},q_{\mbox{\tiny B}}) \hat{\bf T} + A_{\mbox{\tiny N}}(\phi,q_{\mbox{\tiny N}},q_{\mbox{\tiny B}}) \hat{\bf N} + A_{\mbox{\tiny B}}(\phi,q_{\mbox{\tiny N}},q_{\mbox{\tiny B}}) \hat{\bf B}.
\end{equation}
Taking the $q_{\mbox{\tiny N}}$ and $q_{\mbox{\tiny B}} \rightarrow 0$ limit in the gradient operator and performing the dot product with the vector potential yields
\begin{equation}
{\bf A} \cdot \vec{\nabla} = \frac{1}{f(\phi)}A_{\mbox{\tiny T}}(\phi,q_{\mbox{\tiny N}},q_{\mbox{\tiny B}}) \frac{\partial}{\partial \phi} + A_{\mbox{\tiny N}}(\phi,q_{\mbox{\tiny N}},q_{\mbox{\tiny B}}) \frac{\partial}{\partial q_{\mbox{\tiny N}}} + A_{\mbox{\tiny B}}(\phi,q_{\mbox{\tiny N}},q_{\mbox{\tiny B}}) \frac{\partial}{\partial q_{\mbox{\tiny B}}}.
\end{equation}
Finally, the full Hamiltonian is
\begin{align}
\begin{split}
H_\omega^3 =& - \frac{\hbar^2}{2 m_e}\bigg[ \frac{1}{f(\phi)^2} \frac{\partial^2}{\partial \phi^2} -\frac{f'(\phi)}{f(\phi)^3} \frac{\partial}{\partial \phi} -\kappa(\phi) \frac{\partial}{\partial q_{\mbox{\tiny N}}} +\frac{\partial^2}{\partial q_{\mbox{\tiny N}}^2} +\frac{\partial^2}{\partial q_{\mbox{\tiny B}}^2} \\ &+2 i \frac{e}{\hbar} \bigg( \frac{1}{f(\phi)}A_{\mbox{\tiny T}}(\phi,q_{\mbox{\tiny N}},q_{\mbox{\tiny B}}) \frac{\partial}{\partial \phi} + A_{\mbox{\tiny N}}(\phi,q_{\mbox{\tiny N}},q_{\mbox{\tiny B}}) \frac{\partial}{\partial q_{\mbox{\tiny N}}} \\ &+ A_{\mbox{\tiny B}}(\phi,q_{\mbox{\tiny N}},q_{\mbox{\tiny B}}) \frac{\partial}{\partial q_{\mbox{\tiny B}}} \bigg)  -\frac{e^2}{\hbar^2} A(\phi,q_{\mbox{\tiny N}},q_{\mbox{\tiny B}})^2 +V_n(q_{\mbox{\tiny N}})+V_n(q_{\mbox{\tiny B}}) \bigg].
\end{split}
\end{align}
The potentials, $V_n$, normal to the ETH are used to constrain the particle to the helix. The Hamiltonian must now be separated with regard to tangent and normal variables in order to facilitate its reduction to a one-dimensional form.
\section{Constructing the one-dimensional effective Hamiltonian}
To reduce the three dimensional Hamiltonian $H^3_\omega$ to a one dimensional effective Hamiltonian $H^1_\omega$, two procedures must be used. The reduced Laplacian corresponding to this system has been shown, after applying da Costa's reduction procedure, to be \cite{williamson}
\begin{equation}
\nabla^2 \Rightarrow  \frac{1}{f(\phi)^2} \frac{\partial^2}{\partial \phi^2} - \frac{f'(\phi)}{f(\phi)^3} \frac{\partial}{\partial \phi} - 2V_c(\phi) + \frac{\partial^2}{\partial q_{\mbox{\tiny N}}^2} +  \frac{\partial^2}{\partial q_{\mbox{\tiny B}}^2},
\end{equation}
where the geometric curvature potential is
\begin{equation}
V_c(\phi) = -\frac{1}{8} \kappa(\phi)^2.
\end{equation}
In the limit that $q_{\mbox{\tiny N}}$ and $q_{\mbox{\tiny B}} \rightarrow 0$, the form of the ${\bf A} \cdot \vec{\nabla}$ portion of Eq.(18) is not obviously separable into tangent and normal parts. Therefore, any $q_{\mbox{\tiny N}}$ and $q_{\mbox{\tiny B}}$ dependence must be integrated out. The wavefunction is assumed to decouple as the particle is constrained to the ETH, and norm conservation leads to the form $\Psi(\phi,q_{\mbox{\tiny N}},q_{\mbox{\tiny B}}) = \chi_{\mbox{\tiny T}}(\phi) \chi_{\mbox{\tiny N}}(q_{\mbox{\tiny N}}) \chi_{\mbox{\tiny B}}(q_{\mbox{\tiny B}}) G(\phi,q_{\mbox{\tiny N}},q_{\mbox{\tiny B}})^{-1/2}$ \cite{williamson}. Let $I_{\mbox{\tiny N}}$ be the expectation value of $A_{\mbox{\tiny N}}(\phi,q_{\mbox{\tiny N}},q_{\mbox{\tiny B}}) \frac{\partial~~}{\partial q_{\mbox{\tiny N}}}$, and $I_{\mbox{\tiny B}}$ be the expectation value of $A_{\mbox{\tiny B}}(\phi,q_{\mbox{\tiny N}},q_{\mbox{\tiny B}}) \frac{\partial~~}{\partial q_{\mbox{\tiny B}}}$. Then
\begin{align}
\begin{split}
I_{\mbox{\tiny D}} = \int_0^L \hspace{-6pt} &\chi_{\mbox{\tiny D}}^*(q_{\mbox{\tiny D}}) G(\phi,q_{\mbox{\tiny N}})^{-1/2} A_{\mbox{\tiny N}}(\phi,q_{\mbox{\tiny N}},q_{\mbox{\tiny B}}) \\ &\times \big[ \frac{\partial}{\partial q_{\mbox{\tiny D}}} \big( \chi_{\mbox{\tiny D}}(q_{\mbox{\tiny D}}) G(\phi,q_{\mbox{\tiny N}})^{-1/2} \big) \big] G(\phi,q_{\mbox{\tiny N}}) \, dq_{\mbox{\tiny D}},
\end{split}
\end{align}
where ${\rm D}$ is either ${\rm N}$ or ${\rm B}$.
The functions $\chi_{\mbox{\tiny N}}(q_{\mbox{\tiny N}}) G(\phi,q_{\mbox{\tiny N}})^{-1/2}$ and $\chi_{\mbox{\tiny B}}(q_{\mbox{\tiny B}}) G(\phi,q_{\mbox{\tiny N}})^{-1/2}$ can be thought of as the wave functions of a particle in an infinite potential well, although the explicit form of the normal wave function is not essential. Performing the differentiations and integrating by parts allows the non-separable terms to be replaced according to
\begin{equation}
A_{\mbox{\tiny N}}(\phi,q_{\mbox{\tiny N}},q_{\mbox{\tiny B}}) \frac{\partial~~}{\partial q_{\mbox{\tiny N}}} \rightarrow -\frac{1}{2} \frac{\partial~~}{\partial q_{\mbox{\tiny N}}} A_{\mbox{\tiny N}}(\phi,q_{\mbox{\tiny N}},q_{\mbox{\tiny B}}) + \frac{1}{2} \kappa(\phi) A_{\mbox{\tiny N}}(\phi,q_{\mbox{\tiny N}},q_{\mbox{\tiny B}})
\end{equation}
and
\begin{equation}
A_{\mbox{\tiny B}}(\phi,q_{\mbox{\tiny N}},q_{\mbox{\tiny B}}) \frac{\partial~~}{\partial q_{\mbox{\tiny B}}} \rightarrow -\frac{1}{2} \frac{\partial~~}{\partial q_{\mbox{\tiny B}}} A_{\mbox{\tiny B}}(\phi,q_{\mbox{\tiny N}},q_{\mbox{\tiny B}}).
\end{equation}
Derivation of the vector potential using Eq.(15) shows that the explicit form of $A_{\mbox{\tiny N}}(\phi,q_{\mbox{\tiny N}},q_{\mbox{\tiny B}})$ does not depend on $q_{\mbox{\tiny N}}$, and $A_{\mbox{\tiny B}}(\phi,q_{\mbox{\tiny N}},q_{\mbox{\tiny B}})$ does not depend on $q_{\mbox{\tiny B}}$. Therefore
\begin{equation}
\frac{\partial~~}{\partial q_{\mbox{\tiny N}}} A_{\mbox{\tiny N}}(\phi,q_{\mbox{\tiny B}}) = \frac{\partial~~}{\partial q_{\mbox{\tiny B}}} A_{\mbox{\tiny B}}(\phi,q_{\mbox{\tiny N}}) = 0,
\end{equation}
which allows the replacement
\begin{equation}
{\bf A} \cdot \vec{\nabla} \rightarrow \frac{1}{f(\phi)}A_{\mbox{\tiny T}}(\phi,q_{\mbox{\tiny N}},q_{\mbox{\tiny B}}) \frac{\partial}{\partial \phi} + \frac{1}{2} \kappa(\phi) A_{\mbox{\tiny N}}(\phi,q_{\mbox{\tiny B}}).
\end{equation}
Taking the limit as $q_{\mbox{\tiny N}}$ and $q_{\mbox{\tiny B}} \rightarrow 0$ further reduces ${\bf A} \cdot \vec{\nabla}$, leaving only tangential components given by
\begin{equation}
{\bf A} \cdot \vec{\nabla} \rightarrow \frac{1}{f(\phi)}A_{\mbox{\tiny T}}(\phi) \frac{\partial~}{\partial \phi} + \frac{1}{2} \kappa(\phi) A_{\mbox{\tiny N}}(\phi).
\end{equation}
Finally, the ${\bf A}^2$ term of Eq.(4) contains only positive powers of $q_{\mbox{\tiny N}}$ and $q_{\mbox{\tiny B}}$. After taking the $q_{\mbox{\tiny N}}, q_{\mbox{\tiny B}} \rightarrow 0$ limit, only the tangential part of ${\bf A}$ remains, given by,
\begin{equation}
{\bf A} = \hat{\bf T} \big(  B_{\rho} A_{{\mbox{\tiny T}} \rho} + B_z A_{{\mbox{\tiny T}}z} \big) + \hat{\bf N} \big(  B_{\rho} A_{{\mbox{\tiny N}} \rho} + B_z A_{{\mbox{\tiny N}}z} \big) + \hat{\bf B} \big(  B_{\rho} A_{{\mbox{\tiny B}} \rho} + B_z A_{{\mbox{\tiny B}}z} \big).
\end{equation}
Explicit forms of the vector potential components in Eq.(27) are given in the appendix. The subscript $\rho$ in the vector potential terms indicates that term multiplies only the $\rho$-component of the magnetic field. Likewise, the subscript $z$ in the vector potential terms indicates the term multiplies only the $z$-component of the magnetic field.
Now the reduced form of the Hamiltonian (with units) is
\begin{align}
\begin{split}
H^1_\omega =& \frac{1}{2 {\textit m}_e} \bigg[-\hbar^2 \bigg(\frac{1}{f(\phi)^2} \frac{\partial^2}{\partial \phi^2} - \frac{f'(\phi)}{f(\phi)^3} \frac{\partial}{\partial \phi} + \frac{1}{4} \kappa(\phi)^2 \bigg) \\ &- 2 {\textit i} e \hbar \bigg( \frac{1}{f(\phi)}A_{\mbox{\tiny T}}(\phi) \frac{\partial}{\partial \phi} + \frac{1}{2} \kappa(\phi) A_{\mbox{\tiny N}}(\phi) \bigg) + e^2 A(\phi)^2 \bigg]~.
\end{split}
\end{align}
Notice that replacing ${\bf A} \cdot \vec{\nabla}$  of Eq.(4) with the expression in Eq.(26) leads to a second geometric potential, $V_{mag}(\phi)$, in $H^1_\omega$ due to the magnetic field:
\begin{equation}
V_{mag}(\phi) = {\textit i} \kappa(\phi) A_{\mbox{\tiny N}}(\phi).
\end{equation}
The reduced Schr\"{o}dinger equation can be made into a dimensionless form via the following definitions:
\begin{align*}
\gamma_0 =& ~B_z \pi R^2 \\
\gamma_1 =& ~B_{\rho} \pi R^2 \\
\gamma_{\mbox{\tiny N}} =& ~\frac{\pi \hbar}{e} \\
\epsilon ~=& ~\frac{2 {\textit m}_e E R^2}{\hbar^2} \\
\tau_0 =& ~\frac{\gamma_0}{\gamma_{\mbox{\tiny N}}} \\
\tau_1 =& ~\frac{\gamma_1}{\gamma_{\mbox{\tiny N}}},
\end{align*}
where $\tau_0$ and $\tau_1$ should not be confused with the torsion. The dimensionless form of the one-dimensional effective Schr\"odinger equation becomes
\begin{align}
\begin{split}
\bigg( \frac{1}{f(\phi)^2} \frac{\partial^2}{\partial \phi^2} + \bigg[ - \frac{f'(\phi)}{f(\phi)^3} + \frac{2 {\textit i}}{f(\phi)} \big(\tau_1 A_{{\mbox{\tiny T}} \rho} + \tau_0 A_{{\mbox{\tiny T}}z} \big)\bigg] \frac{\partial}{\partial \phi} \\ + \frac{1}{4} \kappa(\phi)^2  + {\textit i} \kappa(\phi) \big(\tau_1 A_{{\mbox{\tiny N}} \rho} + \tau_0 A_{{\mbox{\tiny N}}z} \big) \\
-\big[ \tau_1^2 (A_{{\mbox{\tiny T}} \rho}^2 + A_{{\mbox{\tiny N}} \rho}^2 +A_{{\mbox{\tiny B}} \rho}^2) + \tau_0^2 (A_{{\mbox{\tiny T}}z}^2 + A_{{\mbox{\tiny N}}z}^2 +A_{{\mbox{\tiny B}}z}^2)   \\ + 2 \tau_1 \tau_0 (A_{{\mbox{\tiny T}} \rho} A_{{\mbox{\tiny T}}z} + A_{{\mbox{\tiny N}} \rho} A_{{\mbox{\tiny N}}z} + A_{{\mbox{\tiny B}} \rho} A_{{\mbox{\tiny B}}z}) \big] + \epsilon \bigg) \Psi &= 0.
\end{split}
\end{align}
\section{Computational scheme}
The periodic curvature potentials in the Hamiltonian indicate that the particle's wave function should obey Bloch's theorem. For the $\alpha^{\rm th}$ eigenvalue, a suitable basis set representing single-valuedness and $\omega$-fold symmetry of the system is \cite{williamson}
\begin{equation}
\chi^{p\alpha}(\phi)=\frac{{\rm exp}[i p \phi]}{f(\phi)^{1 / 2}}\sum_n C^{p\alpha}_n {\rm exp} [i n \omega \phi].
\end{equation}
From the ansatz given in Eq.(31), the eigenvalues and eigenvectors of the system were calculated by diagonalizing the $n \times n$ matrix with elements $\langle \chi_m | H^1_\omega | \chi_n \rangle$, or explicitly
\begin{align}
\begin{split}
H_{mn} =& ~\frac{1}{2 \pi} \int_0^{2 \pi} e^{i\omega(n-m)\phi} \bigg( \frac{5}{4} \frac{f'(\phi)^2}{f(\phi)^4} -\frac{f''(\phi)}{2 f(\phi)^3} -2 i (p+ \omega n) \frac{f'(\phi)}{f(\phi)^3} \\ &+ \frac{1}{4} \kappa(\phi)^2  -\frac{(p + \omega n)^2}{f(\phi)^2} - \frac{2 (p + \omega n)}{f(\phi)} (\tau_1 A_{{\mbox{\tiny T}} \rho} + \tau_0 A_{{\mbox{\tiny T}}z}) \\ &- i \frac{f'(\phi)}{f(\phi)^2}(\tau_1 A_{{\mbox{\tiny T}} \rho} + \tau_0 A_{{\mbox{\tiny T}}z}) + i \kappa(\phi)(\tau_1 A_{{\mbox{\tiny N}} \rho} + \tau_0 A_{{\mbox{\tiny N}}z}) \\ &- \big[ \tau_1^2 (A_{{\mbox{\tiny T}} \rho}^2 + A_{{\mbox{\tiny N}} \rho}^2 +A_{{\mbox{\tiny B}} \rho}^2) + \tau_0^2 (A_{{\mbox{\tiny T}}z}^2 + A_{{\mbox{\tiny N}}z}^2 +A_{{\mbox{\tiny B}}z}^2)   \\ &+ 2 \tau_1 \tau_0 (A_{{\mbox{\tiny T}} \rho} A_{{\mbox{\tiny T}}z} + A_{{\mbox{\tiny N}} \rho} A_{{\mbox{\tiny N}}z} + A_{{\mbox{\tiny B}} \rho} A_{{\mbox{\tiny B}z}}) \big] \bigg) \,d\phi.
\end{split}
\end{align}
The basis expansion was limited to five terms ($n \in [-2,2]$) because a $5 \times 5$ matrix proved sufficient to reproduce the the classical toroidal moment calculated using \cite{williamson}
\begin{equation}
\textbf{T}^p_M=-\frac{\pi \omega I a b R}{2} ~\hat{{\bf{k}}},
\end{equation}
with current
\begin{equation}
I=\frac{2  \pi  e \hbar p }{ m_e L^2},
\end{equation}
and total arc length
\begin{equation}
L=\int_0^{2\pi} f(\phi) \,d\phi.
\end{equation}
The eigenvectors of the matrix generated by Eq.(32) allow the calculation of the quantum mechanical current of the $p$-$\alpha$ state, given by \cite{williamson}
\begin{align}
\begin{split}
{\textbf{\textit{j}}}^{p\alpha}(\phi) =& \frac{e \hbar}{m_e} \frac{1}{2\pi} \sum\limits_{m,n} C^{p\alpha}_m C^{p\alpha}_n \bigg[ \frac{(p+\omega n)}{f(\phi)^2} {\rm cos}[\omega(n-m)\phi] \\ &- \frac{f'(\phi)}{2f(\phi)^3} {\rm sin}[\omega(n-m)\phi]\bigg]~\hat{{\bf{T}}},
\end{split}
\end{align}
and the toroidal moments are calculated with \cite{marinov}
\begin{equation}
\textbf{T}^{p\alpha}_M=\frac{1}{10}\int_0^{2\pi} \big[ \big( {\textbf{\textit{j}}}^{p\alpha}(\phi) \cdot {\bf{r}}(\phi) \big) {\bf{r}}(\phi) -2r^2 {\textbf{\textit{j}}}^{p\alpha}(\phi) \big]f(\phi)\,d\phi.
\end{equation}
\section{Results}
Toroidal moments for 4-turn and 8-turn circular and elliptic toroidal helices were calculated using the methods presented in the previous section. The Bloch index was limited to low values, $p=0,1,2$, where the helix is most affected by the magnetic flux. 

The toroidal moment as a function of polar angle $\theta$ (measured from the z-axis) was calculated for circular and elliptic configurations. The magnetic field was swept from $\theta=0$ to $\theta=2 \pi$ radians, while the magnitude of the magnetic field was held constant, i.e., $B_z=B_{max} {\rm cos}(\theta)$ and $B_{\rho}=B_{max} {\rm sin}(\theta)$. Each Bloch index has substate TM's corresponding to the $\alpha^{\rm th}$ energy eigenvalue. As anticipated, for $p=0$, the TM's of the circular toroidal helix quickly drop to zero at $\theta = \frac{\pi}{2}$ and $\theta = \frac{3\pi}{2}$. The toroidal moment magnitude for the 8-turn toroidal helix is larger than that of the 4-turn toroidal helix, but the dependence is not linear in $\omega$, as shown in Fig.1. The elliptic cases also switch sign at the same angles, as shown in Fig.2. In the $p=1$ case, the TM's of the circular toroidal helix reach their maximum magnitude as the magnetic field approaches $\theta= \pi$, as shown in Fig.3. Note the slight assymetry of the toroidal moment due to the lack of symmetry of the TH about the toroidal plane. The $p=1,2$ circular configurations have stiff toroidal moments, showing minimal sensitivity to magnetic field orientation. The elliptic cases also show interesting behavior near $\theta= \pi$ (the TM reaches an extremum) as seen in Fig.4. The TM's remain approximately constant for the $\omega=8$ circular configuration (Figs.5,6). The tall ETH with $\omega=4$ configuration also shows TM sign switching near $\theta = \frac{\pi}{2}$ and $\theta = \frac{3\pi}{2}$. The $\omega=8$ configuration of the tall ETH does not show sign switching, but does show a decrease in TM magnitude near $\theta= \pi$. The flat ETH shows TM sign switching in the $\omega=4$ configuration, but only the TM corresponding to the lowest eigenvalue switches sign in the $\omega=8$ configuration (Fig.6). The toroidal moment was independent of the magnetic field's azimuthal orientation. The elliptic configurations are more sensitive to the magnetic field orientation because the field ``sees" a larger effective radius, corresponding to larger magnetic flux.

The dependence of the toroidal moments on magnetic field magnitude was also investigated. The $\rho$ and $z$ directions were treated independently in order to distinguish the effects of the magnetic field's vertical and horizontal flux through the toroidal helix. The ratio of vertical magnetic flux to the magnetic flux quanta was represented by $\tau_0$, and the ratio of the horizontal magnetic flux to the magnetic flux quanta was represented by $\tau_1$, as defined in the previous section. For $p=0$, the TM's corresponding to the lowest energy eigenvalue in the circular configuration remained near zero as $\tau_0$ increased, while the higher energy TM's jumped from zero to approximately constant non-zero values (Fig.7), roughly independent of $\omega$. The circular configuration TM's reached saturation almost as soon as the flux became nonzero. The TM's of the elliptic cases changed magnitude much more gradually (Fig.8). 

There was no induced TM as $\tau_1$ was increased from zero in the $p=0$ case. For $p=1$, the TM's due to $\tau_1$ are approximately constant for all energies as $\tau_1$ increases. The TM's due to $\tau_1$ also remain constant as $\tau_1$ increases. This behavior holds for both the $\omega=4$ and $\omega=8$ configurations. The $p=2$ toroidal moment behavior is similar to the $p=1$ case. Therefore, the toroidal moment as a function of $\tau_1$ is not presented graphically.

Finally, the dependence of the toroidal moment on eccentricity was investigated. The toroidal moment was calculated as a function of minor vertical, and minor horizontal axes. First, $a$ was held constant while $b$, the minor vertical axis, was increased incrementally from $0.1$ to $0.9$ with $\tau_0$ held constant at $\tau_0=2$. The toroidal moment was also calculated as $b$ was held constant along with $\tau_0$, while $a$, the minor horizontal axis, was increased incrementally from $0.1$ to $0.9$. The magnitude of the toroidal moment corresponding to each current state quickly attained a saturation value and remained near that value as the eccentricity was increased. The $\omega=8$ cases were more stable than their $\omega=4$ counterparts (Fig.9 and Fig.10.).

\section{Conclusions}
This work has demonstrated that curvature terms are essential for a proper description of a particle constrained to a one-dimensional manifold. The effect of the curvature potential $V_c(\phi)$, which does not couple to the vector potential, has been discussed in previous work \cite{williamson}. A second potential, $V_{mag}(\phi)$, due to the curvature of the ETH coupling to the vector potential, appeared when the full Hamiltonian was reduced to a one-dimensional effective Hamiltonian. Omission of this magnetic curvature potential leads to a non-Hermitian Hamiltonian matrix; when included, a Hermitian matrix results.

Toroidal moments were shown to increase or diminish, depending both on the magnitude and orientation of the magnetic field, and the configuration of the toroidal helix system. The most striking effect was the dependence of the toroidal moment on the magnetic field's polar orientation. The toroidal moment's drop to zero at $\theta= \pi/2$ and $\theta=3 \pi/2$ in the $p=0$ circular configuration, and the sign switching of the toroidal moment at (or very near to) those angles in the elliptic $p=0$ configurations, could possibly be exploited in experiments or devices with toroidal moment sensitivity. The effects of the curvature of toroidal helices on the quantum mechanical current states, and their corresponding toroidal moments may be important for calculating photon emission due to an external electromagnetic field\cite{agafonov}. The inclusion of spin in this one-dimensional model is also of interest. The authors intend to investigate these issues in later work.
\newpage
\renewcommand{\theequation}{A-\arabic{equation}}
\setcounter{equation}{0}  
\allowdisplaybreaks
\section*{Appendix}  

Explicit forms of the vector potential components are found by expanding the Cartesian unit vectors, $\hat{{\rm i}}, \hat{{\rm j}}, \hat{{\rm k}},$ and the cylindrical unit vector $\hat{ \phi}$ in the Frenet basis, (e.g. $\hat{{\rm i}} = {\rm i}_T \hat{T} + {\rm i}_N \hat{N} + {\rm i}_B \hat{B}$).

\vspace{8pt}
\noindent The following definitions,
\begin{align}
\begin{split}
W(\phi) =& ~R+a~{\rm cos}(\omega\phi)
\end{split}\\
\begin{split}
P(\phi) =& ~\big( a^2 {\rm sin}^2(\omega \phi) + b^2 {\rm cos}^2(\omega \phi) \big)^{1/2}
\end{split}\\
\begin{split}
f(\phi) =& ~\big( \omega^2 P(\phi)^2 + W(\phi)^2 \big)^{1/2}
\end{split}\\
\begin{split}
P_1(\phi) =& -\frac{b}{f(\phi)^2 P(\phi)}~ \big( a~\omega^2 + W(\phi) {\rm cos}(\omega \phi) \big)
\end{split}\\
\begin{split}
P_2(\phi) =& ~\frac{{\rm sin}(\omega \phi)}{f(\phi) P(\phi)} \bigg(a + \frac{\omega^2 W(\phi) (a^2 - b^2) {\rm cos}(\omega \phi) + P(\phi)^2 a~\omega^2}{f(\phi)^2 } \bigg),
\end{split}
\end{align}
with the curvature
\begin{align}
\kappa (\phi) =& \big( P_1(\phi)^2 + P_2(\phi)^2 \big)^{1/2}, \hspace{5cm}
\end{align}
and the unit vector components
\begin{align}
\begin{split}
{\rm i}_{\mbox{\tiny T}}(\phi) =& - \frac{1}{f(\phi)} \big( a~\omega~{\rm sin}(\omega \phi) {\rm cos}(\phi) + W(\phi) {\rm sin}(\phi) \big)
\end{split}\\
\begin{split}
{\rm i}_{\mbox{\tiny N}}(\phi) =& \frac{1}{\kappa(\phi) P(\phi)} P_1(\phi)~b~{\rm cos}(\omega \phi) {\rm cos}(\phi) \\ &- \frac{1}{\kappa(\phi) P(\phi) f(\phi)} P_2(\phi)~a~W(\phi) {\rm sin}(\omega \phi) {\rm cos}(\phi) \\ &+ \frac{1}{\kappa(\phi) f(\phi)} P_2(\phi) P(\phi)~\omega~{\rm sin}(\phi)
\end{split}\\
\begin{split}
{\rm i}_{\mbox{\tiny B}}(\phi) =& \frac{1}{\kappa(\phi) P(\phi)} P_2(\phi)~b~{\rm cos}(\omega \phi) {\rm cos}(\phi) \\ &+ \frac{1}{\kappa(\phi) P(\phi) f(\phi)} P_1(\phi)~a~W(\phi) {\rm sin}(\omega \phi) {\rm cos}(\phi) \\ &- \frac{1}{\kappa(\phi) f(\phi)} P_1(\phi) P(\phi)~\omega~{\rm sin}(\phi)
\end{split}\\
\begin{split}
{\rm j}_{\mbox{\tiny T}}(\phi) =& \frac{1}{f(\phi)} \big(-a~\omega ~ {\rm sin}(\omega \phi) {\rm sin}(\phi) + W(\phi) {\rm cos}(\phi)\big)
\end{split}\\
\begin{split}
{\rm j}_{\mbox{\tiny N}}(\phi) =& \frac{1}{\kappa(\phi) P(\phi)} P_1(\phi)~b~{\rm cos}(\omega \phi) {\rm sin}(\phi) \\ &- \frac{1}{\kappa(\phi) P(\phi) f(\phi)} P_2(\phi)~a~W(\phi) {\rm sin}(\omega \phi) {\rm sin}(\phi) \\ &- \frac{1}{\kappa(\phi) f(\phi)} P_2(\phi) P(\phi)~\omega~{\rm cos}(\phi)
\end{split}\\
\begin{split}
{\rm j}_{\mbox{\tiny B}}(\phi) =& \frac{1}{\kappa(\phi) P(\phi)} P_2(\phi)~b~{\rm cos}(\omega \phi) {\rm sin}(\phi) \\ &+ \frac{1}{\kappa(\phi) P(\phi) f(\phi)} P_1(\phi)~a~W(\phi) {\rm sin}(\omega \phi) {\rm sin}(\phi) \\ &+ \frac{1}{\kappa(\phi) f(\phi)} P_1(\phi) P(\phi)~\omega~{\rm cos}(\phi)
\end{split}\\
\begin{split}
{\rm k}_{\mbox{\tiny T}}(\phi) =& \frac{1}{f(\phi)} b~\omega~{\rm cos}(\omega \phi)
\end{split}\\
\begin{split}
{\rm k}_{\mbox{\tiny N}}(\phi) =& \frac{1}{\kappa(\phi) P(\phi)} \bigg( P_1(\phi)~a~{\rm sin}(\omega \phi) + \frac{P_2(\phi)~b~W(\phi) {\rm cos}(\omega \phi)}{f(\phi)} \bigg)
\end{split}\\
\begin{split}
{\rm k}_{\mbox{\tiny B}}(\phi) =& \frac{1}{\kappa(\phi) P(\phi)} \bigg( P_2(\phi)~a~{\rm sin}(\omega \phi) - \frac{P_1(\phi)~b~W(\phi) {\rm cos}(\omega \phi)}{f(\phi)} \bigg)
\end{split}\\
\begin{split}
\phi_{\mbox{\tiny T}}(\phi) =& W(\phi)/f(\phi)
\end{split}\\
\begin{split}
\phi_{\mbox{\tiny N}}(\phi) =& - \frac{\omega}{\kappa(\phi) f(\phi)} P_2(\phi) P(\phi)
\end{split}\\
\begin{split}
\phi_{\mbox{\tiny B}}(\phi) =& \frac{\omega}{\kappa(\phi) f(\phi)} P_1(\phi) P(\phi),
\end{split}
\end{align}
allow the components of the vector potential to be written
\begin{align}
\begin{split}
A_{{\mbox{\tiny T}} \rho}(\phi) =& ~\frac{1}{2} W(\phi) {\rm k}_{\mbox{\tiny T}}(\phi) \big( {\rm cos}(\phi_{\mbox{\tiny M}}){\rm sin}(\phi) - {\rm sin}(\phi_{\mbox{\tiny M}}) {\rm cos}(\phi) \big) \\ &+ \frac{1}{2}~b~{\rm sin}(\omega \phi) {\rm sin}(\phi_{\mbox{\tiny M}}) {\rm i}_{\mbox{\tiny T}}(\phi)  - \frac{1}{2}~b~{\rm sin}(\omega \phi) {\rm cos}(\phi_{\mbox{\tiny M}}) {\rm j}_{\mbox{\tiny T}}(\phi)
\end{split}\\
A_{{\mbox{\tiny T}}z}(\phi) =& ~\frac{1}{2} W(\phi) \phi_{\mbox{\tiny T}}(\phi) \\
\begin{split}
A_{{\mbox{\tiny N}} \rho}(\phi) =& ~\frac{1}{2} W(\phi) {\rm k}_{\mbox{\tiny N}}(\phi) \big( {\rm cos}(\phi_{\mbox{\tiny M}}) {\rm sin}(\phi) - {\rm sin}(\phi_{\mbox{\tiny M}}) {\rm cos}(\phi) \big) \\ &+ \frac{1}{2}~b~{\rm sin}(\omega \phi) {\rm sin}(\phi_{\mbox{\tiny M}}) {\rm i}_{\mbox{\tiny N}}(\phi) - \frac{1}{2}~b~{\rm sin}(\omega \phi) {\rm cos}(\phi_{\mbox{\tiny M}}) {\rm j}_{\mbox{\tiny N}}(\phi)
\end{split}\\
A_{{\mbox{\tiny N}}z}(\phi) =& ~\frac{1}{2} W(\phi) \phi_{\mbox{\tiny N}}(\phi) \\
\begin{split}
A_{{\mbox{\tiny B}} \rho}(\phi) =& ~\frac{1}{2} W(\phi) {\rm k}_{\mbox{\tiny B}}(\phi) \big( {\rm cos}(\phi_{\mbox{\tiny M}}) {\rm sin}(\phi) - {\rm sin}(\phi_{\mbox{\tiny M}}) {\rm cos}(\phi) \big) \\ &+ \frac{1}{2}~b~{\rm sin}(\omega \phi) {\rm sin}(\phi_{\mbox{\tiny M}}) {\rm i}_{\mbox{\tiny B}}(\phi) - \frac{1}{2}~b~{\rm sin}(\phi_{\mbox{\tiny M}}) {\rm cos}(\phi_{\mbox{\tiny M}}) {\rm j}_{\mbox{\tiny B}}(\phi)
\end{split}\\
A_{{\mbox{\tiny B}z}}(\phi) =& ~\frac{1}{2} W(\phi) \phi_{\mbox{\tiny B}}(\phi).
\end{align}

\newpage
\section*{References}
\bibliography{refs2comp}

\begin{thebibliography}{10}
\expandafter\ifx\csname url\endcsname\relax
  \def\url#1{\texttt{#1}}\fi
\expandafter\ifx\csname urlprefix\endcsname\relax\def\urlprefix{URL }\fi
\expandafter\ifx\csname href\endcsname\relax
  \def\href#1#2{#2} \def\path#1{#1}\fi

\bibitem{kaelscience}
T.~Kaelberer, V.~Fedetov, N.~Papasimakis, D.~Tsai, N.~Zheludev, Science 330
  (2010) 1510.

\bibitem{papas}
N.~Papasimakis, V.~A. Fedotov, K.~Marinov, N.~I. Zheludev, Phys. Rev. Lett. 103
  (2009) 093901.

\bibitem{marinov}
K.~Marinov, A.~D. Boardman, V.~A. Fedotov, N.~Zheludev2, N. J. Phys. 9 (2007)
  324.

\bibitem{williamson}
M.~Encinosa, J.~Williamson, {Toroidal Moments of Schr\"odinger Eigenstates}
  (2011).
\newblock \href {http://arxiv.org/abs/1106.4248v1, Submitted to Physica E}
  {\path{arXiv:1106.4248v1, Submitted to Physica E}}.

\bibitem{afanasiev}
G.~F. Afanasiev, V.~M. Dubovik, G.~Goldoni, F.~Troiani, E.~Molinari, Phys.
  Part. Nucl. 29 (1998) 366.

\bibitem{ceulemans}
A.~Ceulemans, L.~Chibotaru, P.~Fowler, Phys. Rev. Lett. 80 (1998) 1861.

\bibitem{dubovik}
V.~M. Dubovik, V.~V. Tugushev, Phys. Rep. 187 (1990) 145.

\bibitem{naumov}
I.~Naumov, L.~Bellaiche, H.~Fu, Nature 432 (2004) 737.

\bibitem{spaldin}
N.~A. Spaldin, M.~Fiebig, M.~Mostovoy, J. Phys.: Condens. Matter 20 (2008) 1.

\bibitem{sawada}
K.~Sawada, N.~Nagaosa, Phys. Rev. Lett. 95 (2005) 237402.

\bibitem{chapblick}
A.~Chaplik, R.~H. Blick, New J. Phys. 6 (2004) 33.

\bibitem{dacosta1}
R.~C.~T. da~Costa, Phys. Rev. A 23 (1981) 1982.

\bibitem{dacosta2}
R.~C.~T. da~Costa, Phys. Rev. A 25 (1982) 2893.

\bibitem{duclosexner}
P.~Duclos, P.~Exner, Rev. Math. Phys. 7 (1995) 73.

\bibitem{ee1}
M.~Encinosa, B.~Etemadi, PRA 58 (1998) 77.

\bibitem{ee2}
M.~Encinosa, B.~Etemadi, Physica B 266 (1998) 361.

\bibitem{jenskoppe}
B.~Jensen, H.~Koppe, Ann. of Phys. 63 (1971) 586.

\bibitem{matsutani1}
S.~Matusani, J. Phys. Soc. Jap. 61 (1991) 55.

\bibitem{matsutani2}
S.~Matsutani, Rev. Math. Phys. 11 (1999) 171.

\bibitem{taira}
H.~Taira, H.~Shima, Surf. Sci. 601 (2007) 5270.

\bibitem{schujaff}
P.~C. Schuster, R.~L. Jaffe, Ann. Phys. 307 (2003) 132.

\bibitem{burgsjens}
M.~Burgess, B.~Jensen, Phys. Rev. A 48 (1993) 1861.

\bibitem{encarbB}
M.~Encinosa, Physica E 28 (2005) 209.

\bibitem{agafonov}
A.~I. Agafonov, {Electromagnetic-field-induced decay of currents in thin-film
  superconducting rings with photons emission} (2011).
\newblock \href {http://arxiv.org/abs/1107.2905v1} {\path{arXiv:1107.2905v1}}.

\end{thebibliography}
\newpage

\begin{figure}[h]
\centering
\subfloat[Subfigure 1a list of figures text][4-turn toroidal helix.]{
\includegraphics[width=0.4\textwidth]{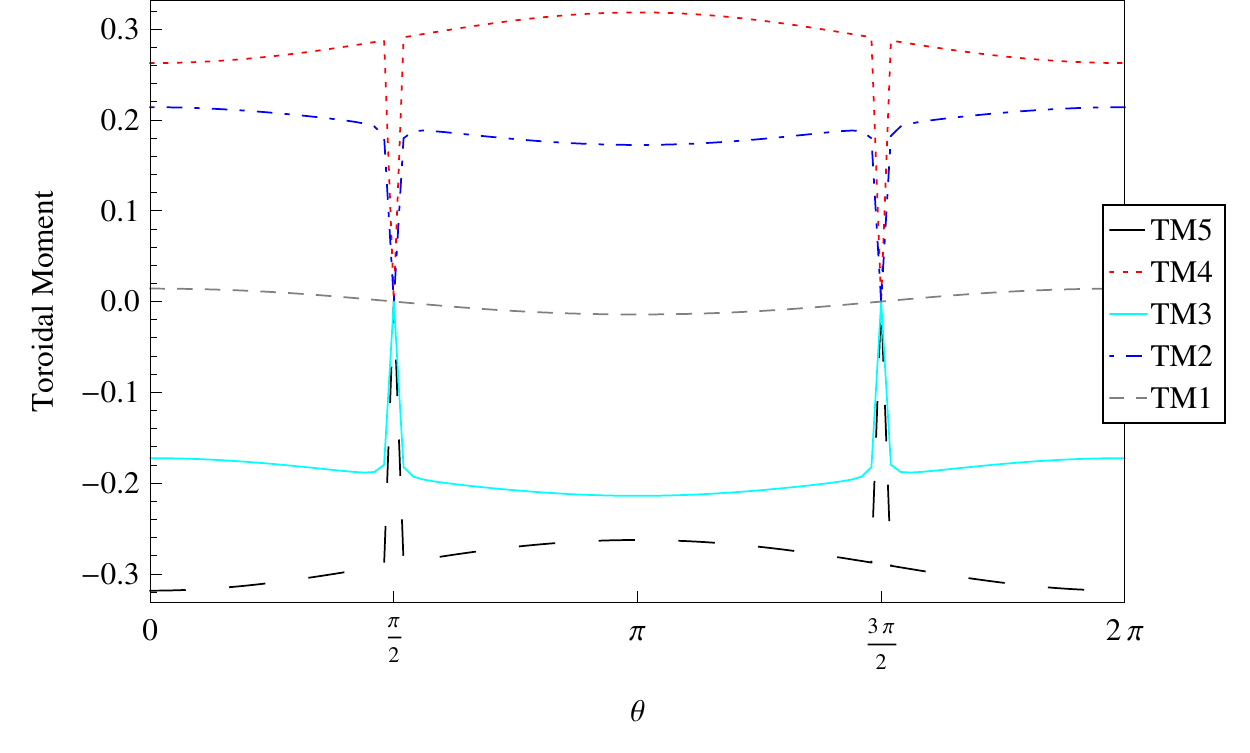}
\label{fig:subfig1a}}
\qquad
\subfloat[Subfigure 1b list of figures text][8-turn toroidal helix.]{
\includegraphics[width=0.4\textwidth]{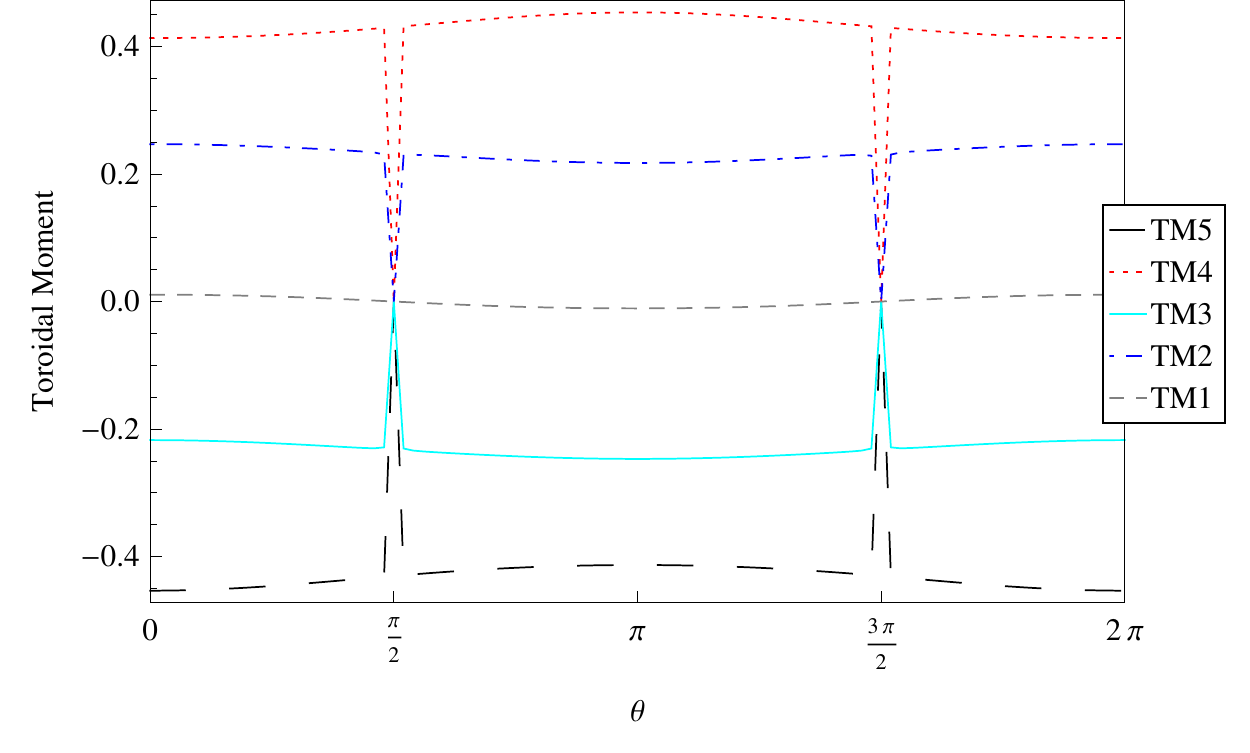}
\label{fig:subfig1b}}
\caption{Toroidal moment (in units of $e \hbar R/m_e $) as a function of polar angle, $\theta$, for the circular toroidal helix configuration $R=1,~a=0.5,~b=0.5,~p=0$.}
\label{fig:globfig1}
\end{figure}

\begin{figure}[h]
\centering
\subfloat[Subfigure 2a list of figures text][4-turn, tall elliptic toroidal helix with $a=0.25$, and $b=0.75$.]{
\includegraphics[width=0.4\textwidth]{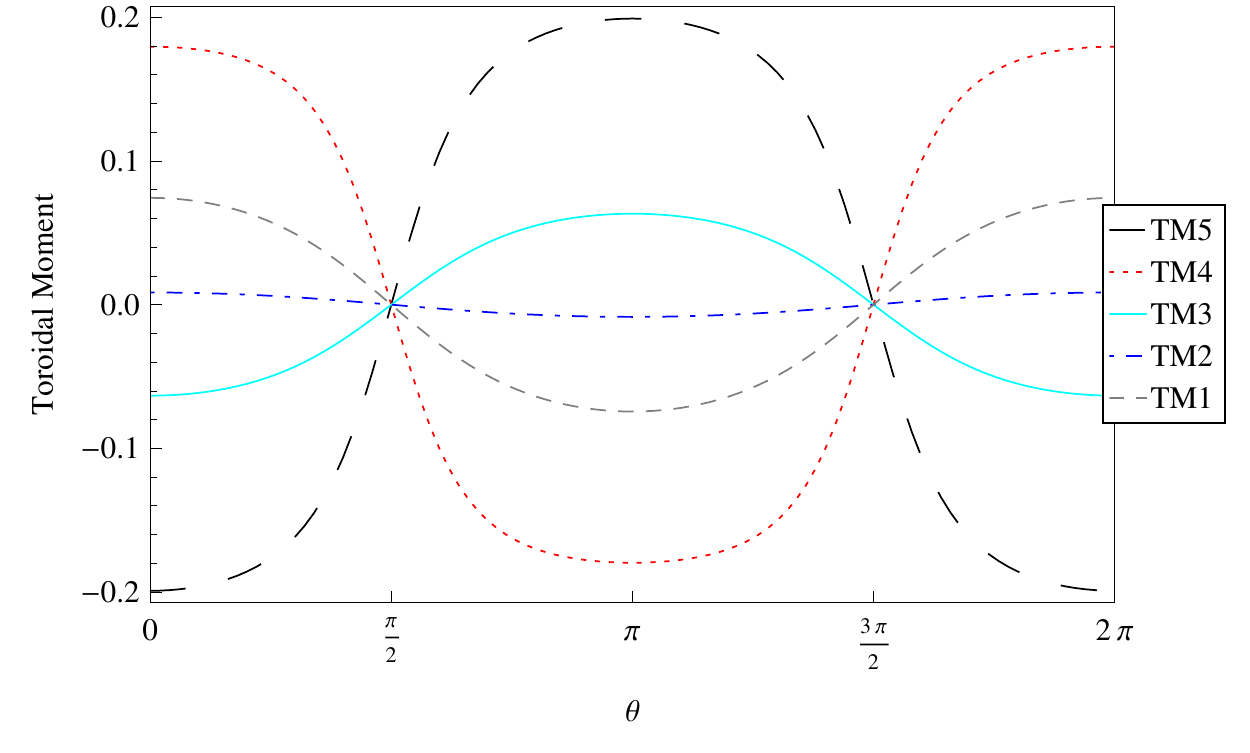}
\label{fig:subfig2a}}
\qquad
\subfloat[Subfigure 2b list of figures text][4-turn, flat elliptic toroidal helix with $a=0.75$, and $b=0.25$.]{
\includegraphics[width=0.4\textwidth]{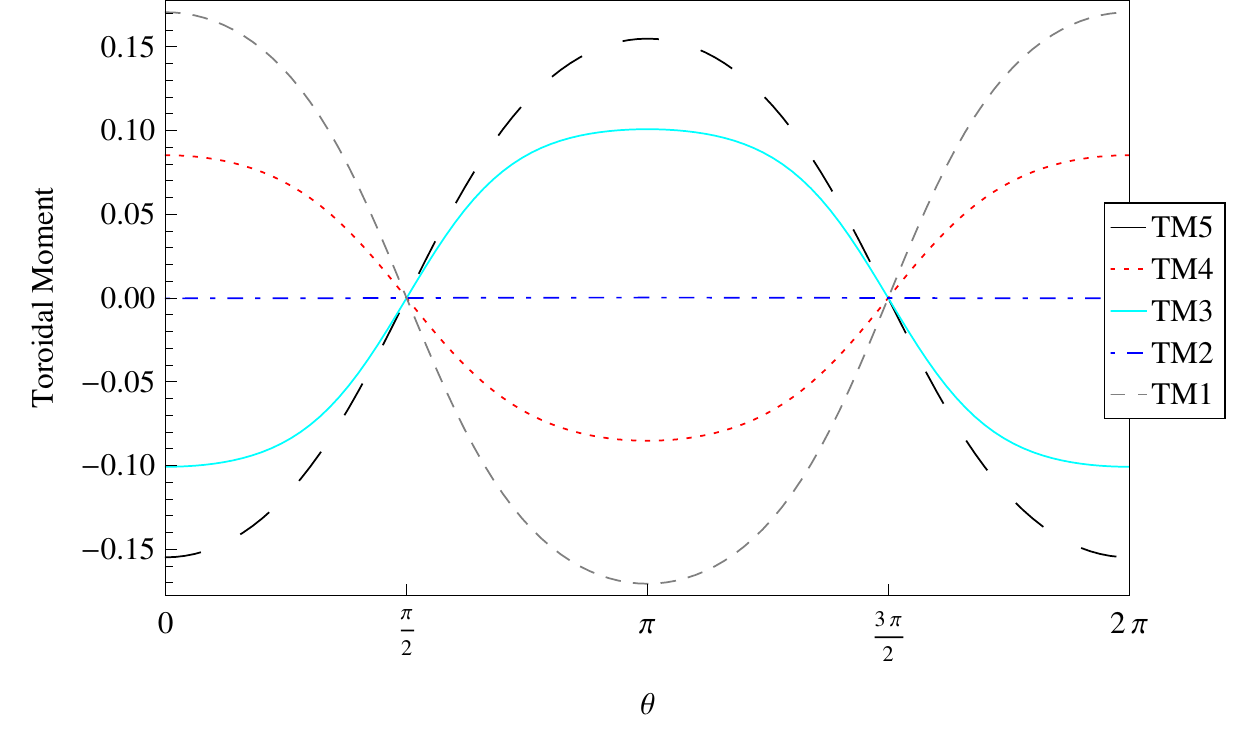}
\label{fig:subfig2b}}
\qquad
\subfloat[Subfigure 2c list of figures text][8-turn, tall elliptic toroidal helix with $a=0.25$, and $b=0.75$.]{
\includegraphics[width=0.4\textwidth]{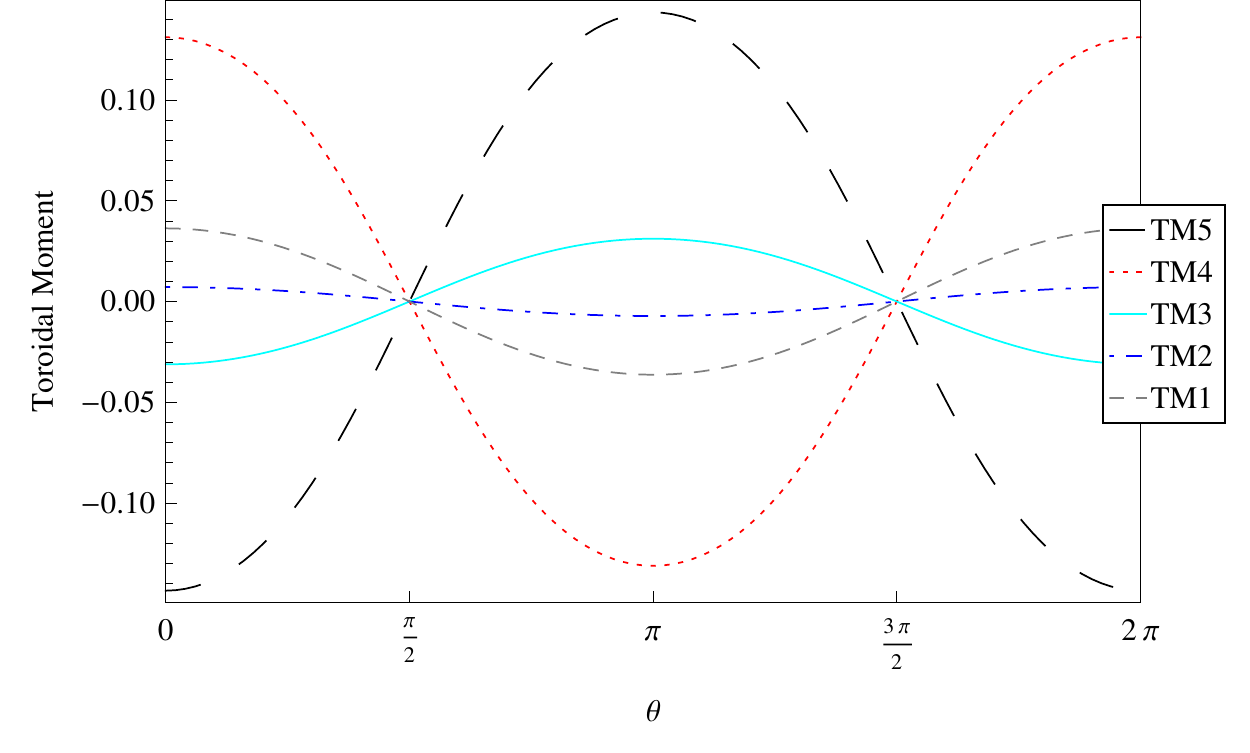}
\label{fig:subfig2c}}
\qquad
\subfloat[Subfigure 2d list of figures text][8-turn, flat elliptic toroidal helix with $a=0.75$, and $b=0.25$.]{
\includegraphics[width=0.4\textwidth]{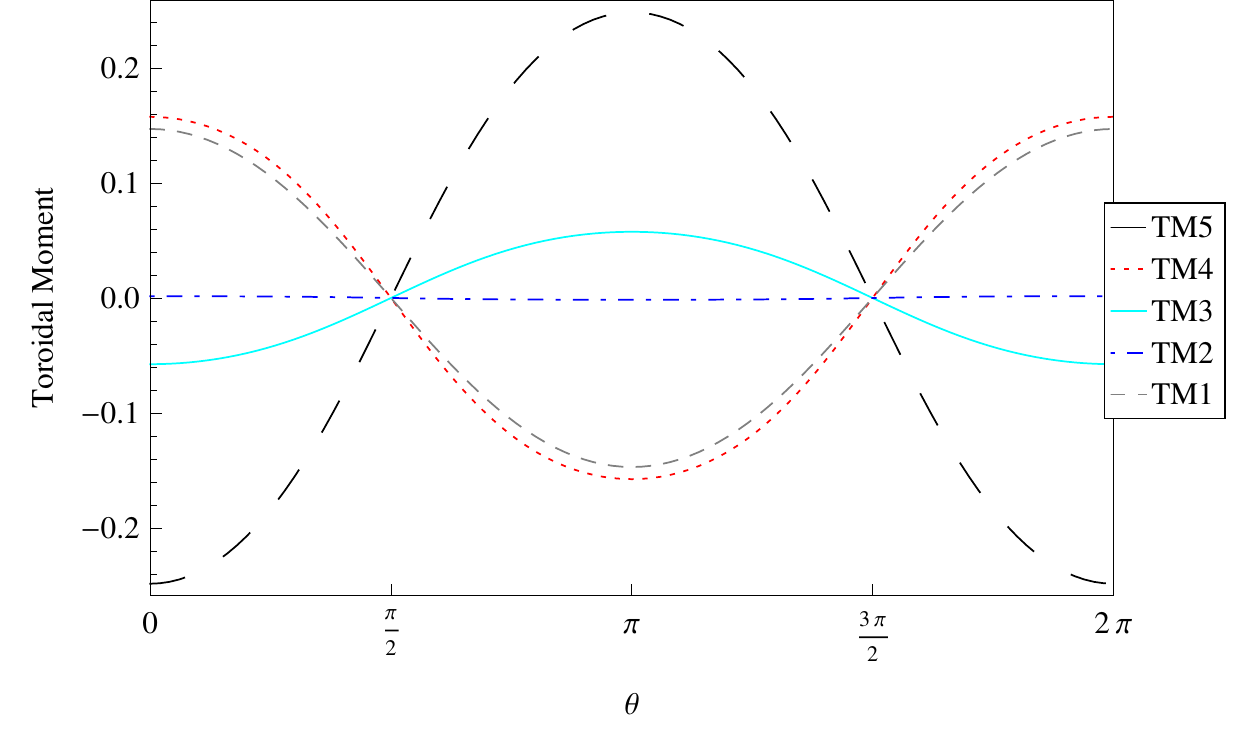}
\label{fig:subfig2d}}
\caption{Toroidal moments (in units of $e \hbar R/m_e $) as a function of $\theta$ for the $p=0$ elliptic configurations.}
\label{fig:globfig2}
\end{figure}

\begin{figure}[h]
\centering
\subfloat[Subfigure 3a list of figures text][4-turn, circular toroidal helix with $a=0.5$, and $b=0.5$.]{
\includegraphics[width=0.4\textwidth]{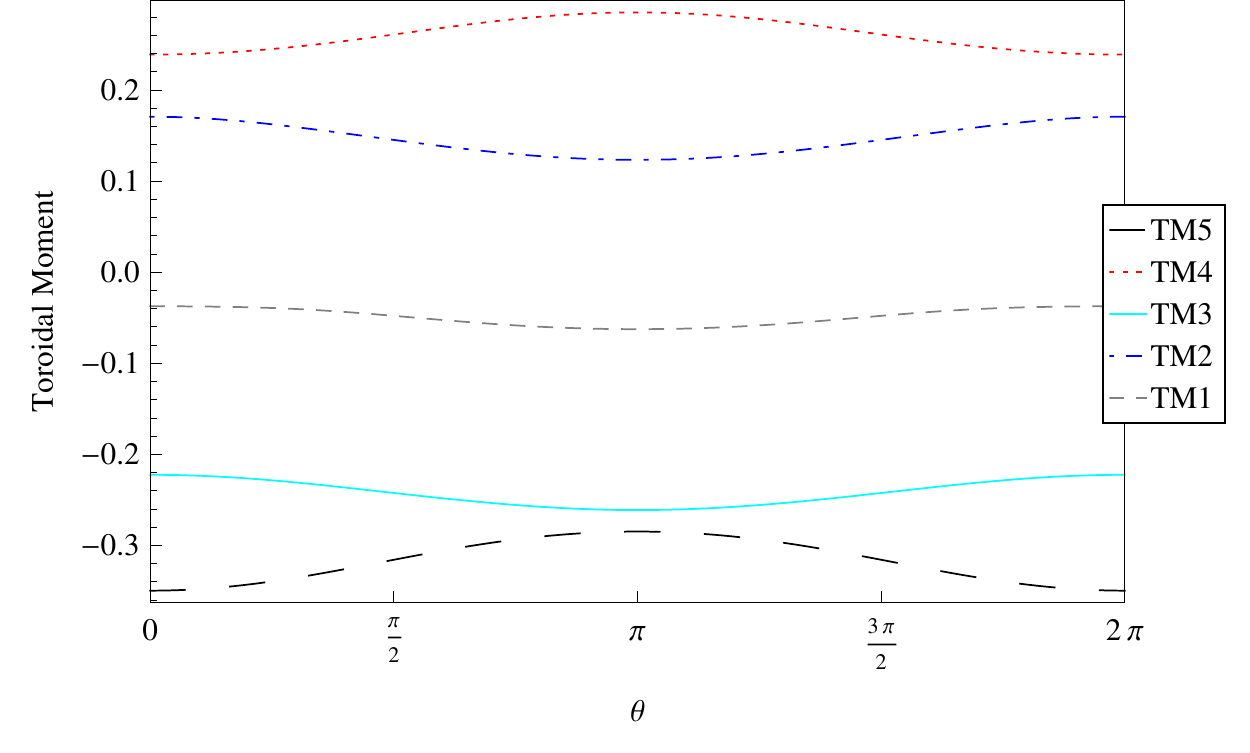}
\label{fig:subfig3a}}
\qquad
\subfloat[Subfigure 3b list of figures text][8-turn, circular toroidal helix with $a=0.5$, and $b=0.5$.]{
\includegraphics[width=0.4\textwidth]{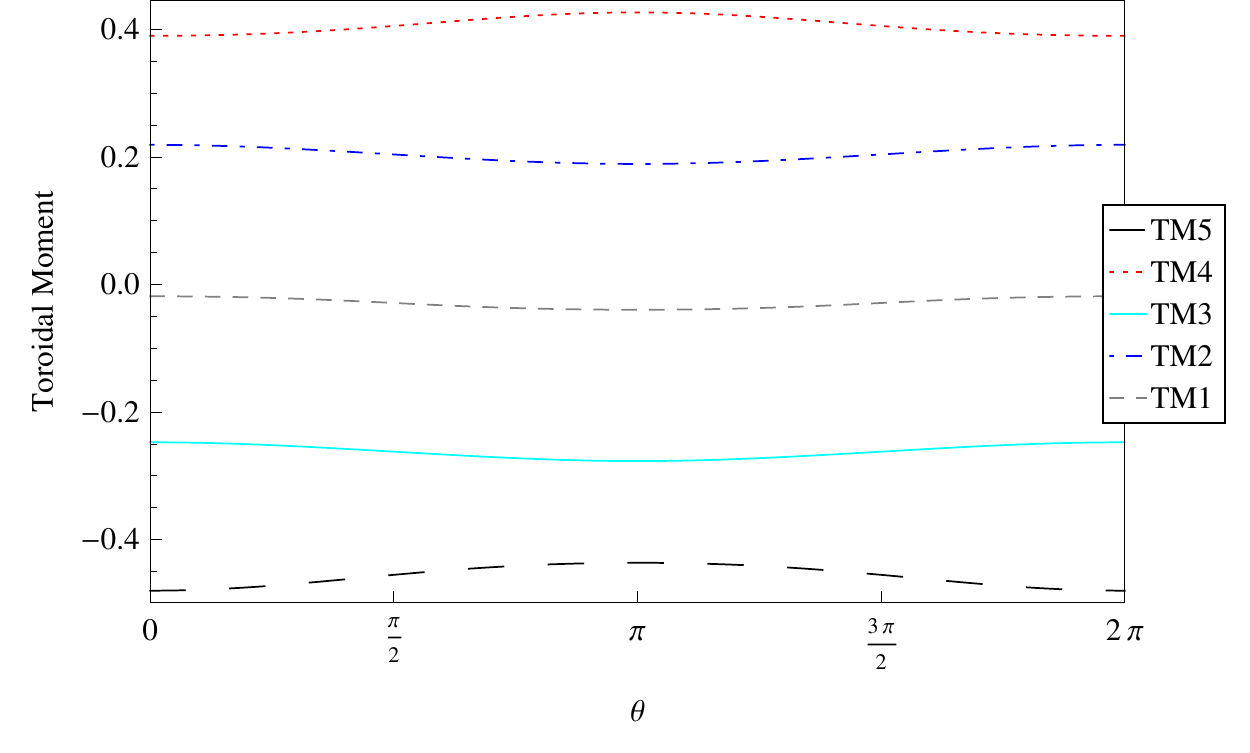}
\label{fig:subfig3b}}
\qquad
\caption{The toroidal moments (in units of $e \hbar R/m_e $) as a function of $\theta$, for the $p=1$ circular cconfigurations reach their maximum magnitude as the magnetic field (held at constant magnitude) approaches $\theta= \pi$.}
\label{fig:globfig3}
\end{figure}

\begin{figure}[h]
\centering
\subfloat[Subfigure 4a list of figures text][4-turn, tall elliptic toroidal helix with $a=0.25$, and $b=0.75$.]{
\includegraphics[width=0.4\textwidth]{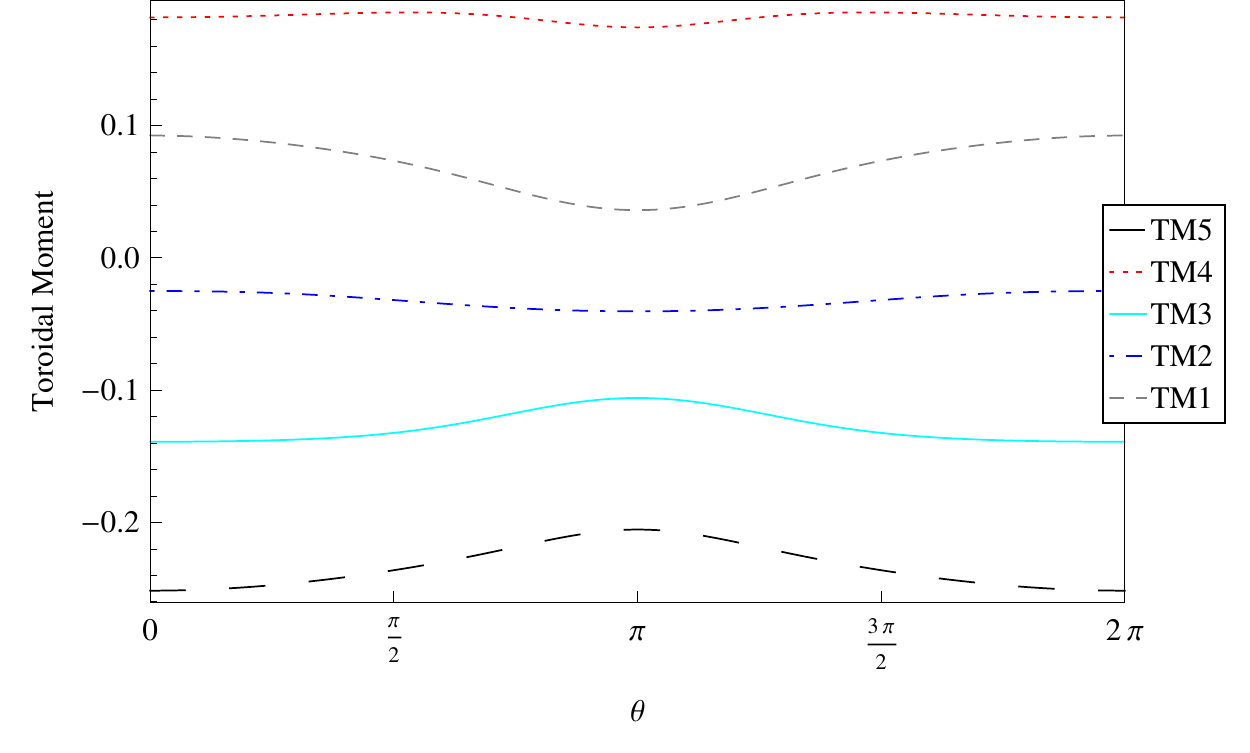}
\label{fig:subfig4a}}
\qquad
\subfloat[Subfigure 4b list of figures text][4-turn, flat elliptic toroidal helix with $a=0.75$, and $b=0.25$.]{
\includegraphics[width=0.4\textwidth]{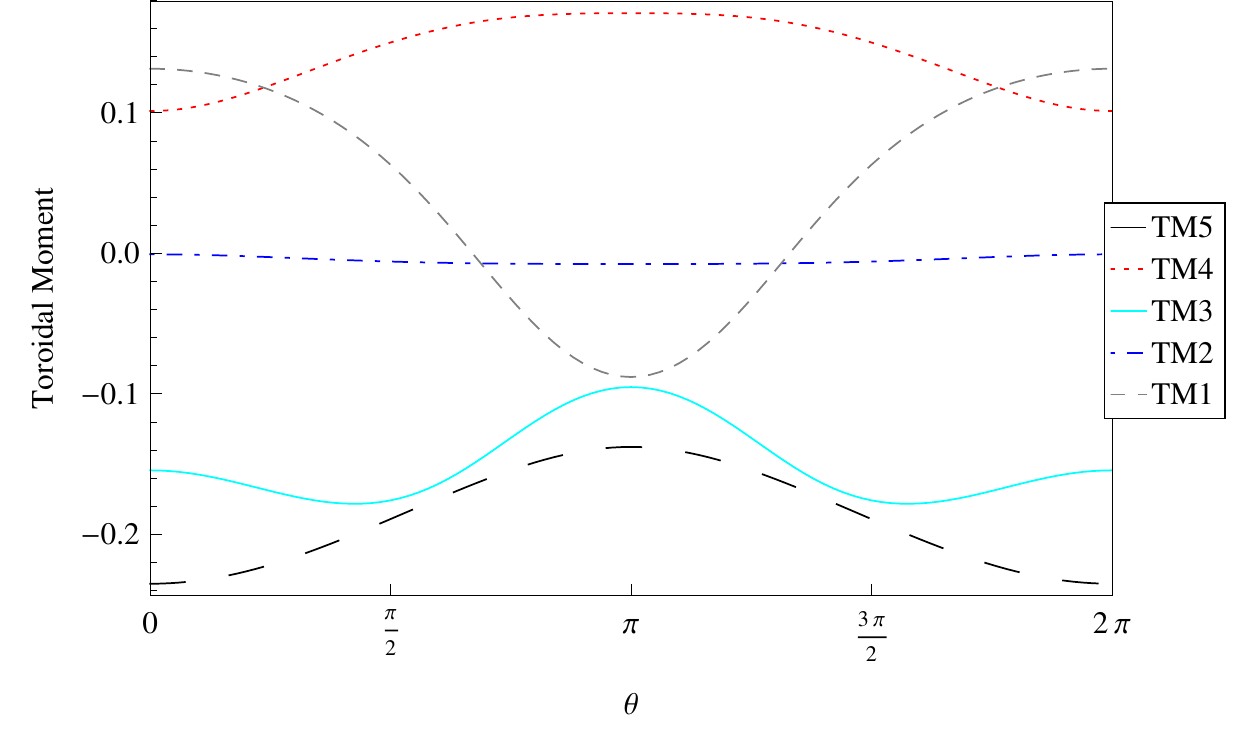}
\label{fig:subfig4b}}
\qquad
\subfloat[Subfigure 4c list of figures text][8-turn, tall elliptic toroidal helix with $a=0.25$, and $b=0.75$.]{
\includegraphics[width=0.4\textwidth]{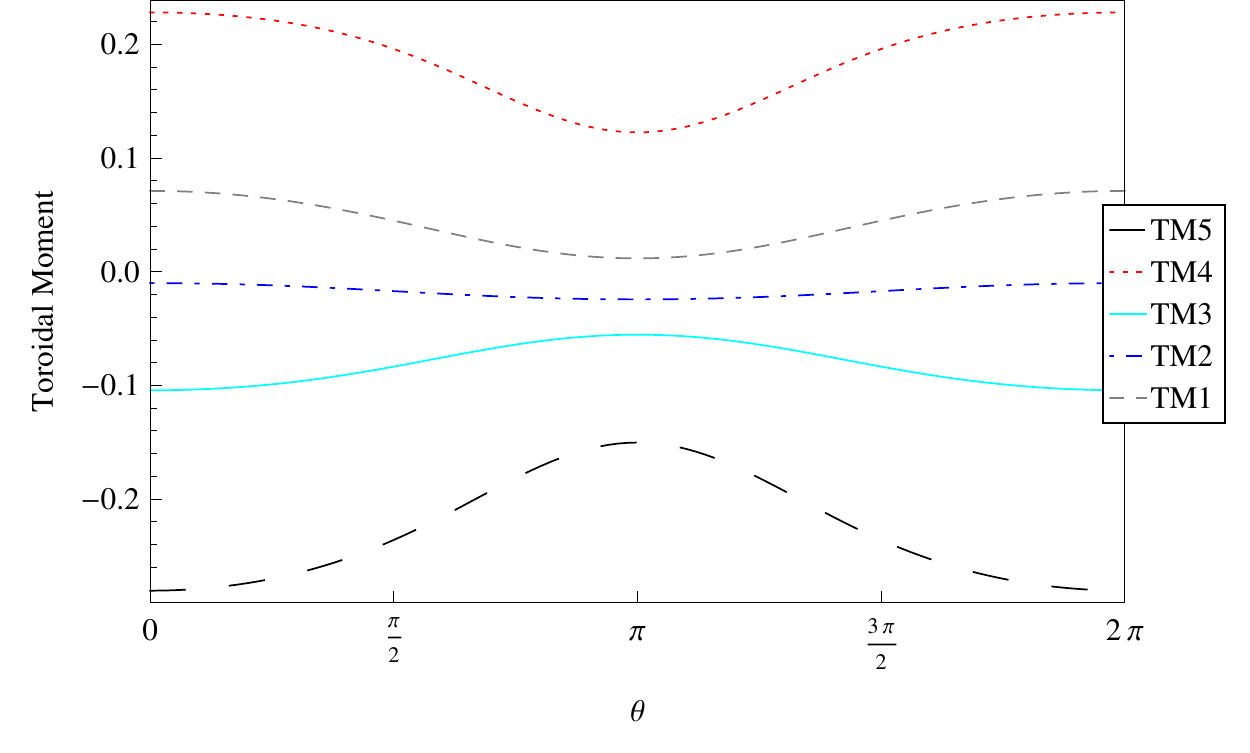}
\label{fig:subfig4c}}
\qquad
\subfloat[Subfigure 4d list of figures text][8-turn, flat elliptic toroidal helix with $a=0.75$, and $b=0.25$.]{
\includegraphics[width=0.4\textwidth]{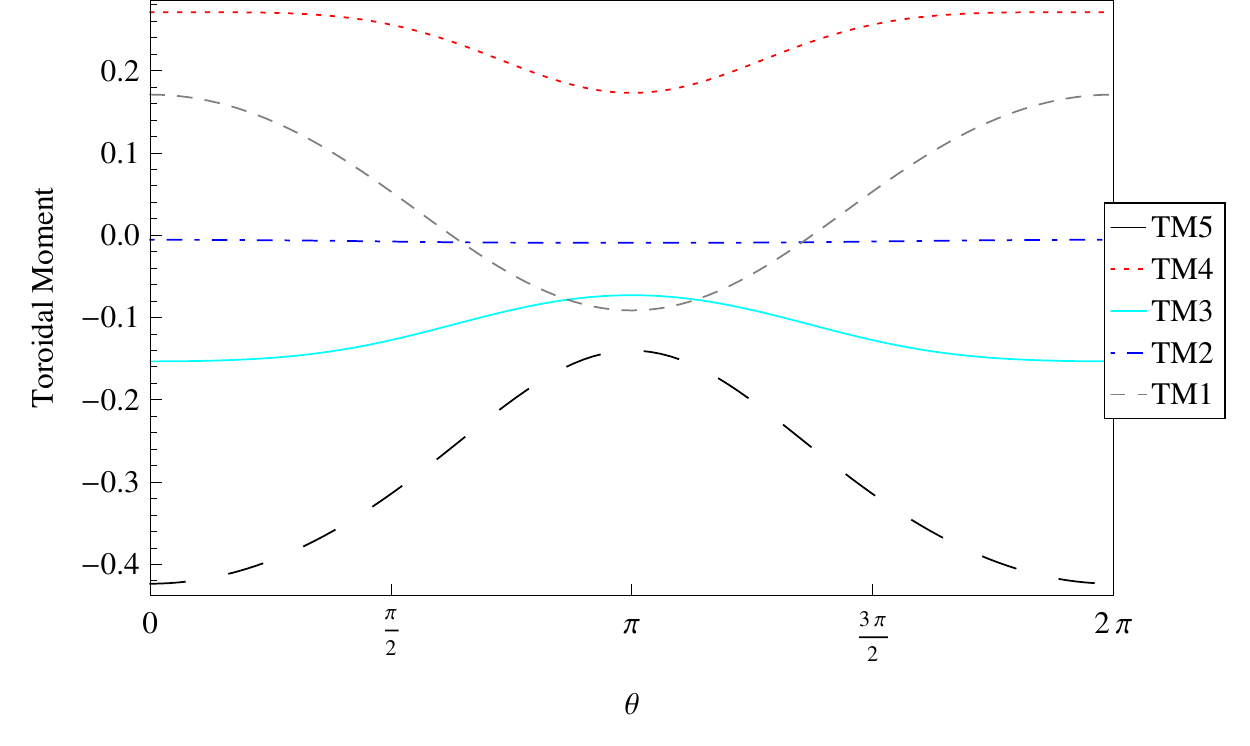}
\label{fig:subfig4d}}
\caption{Toroidal moments (in units of $e \hbar R/m_e $) as a function of $\theta$ for $p=1$.}
\label{fig:globfig4}
\end{figure}

\begin{figure}[h]
\centering
\subfloat[Subfigure 5a list of figures text][4-turn, circular toroidal helix with $R=1$, $a=0.5$, and $b=0.5$.]{
\includegraphics[width=0.4\textwidth]{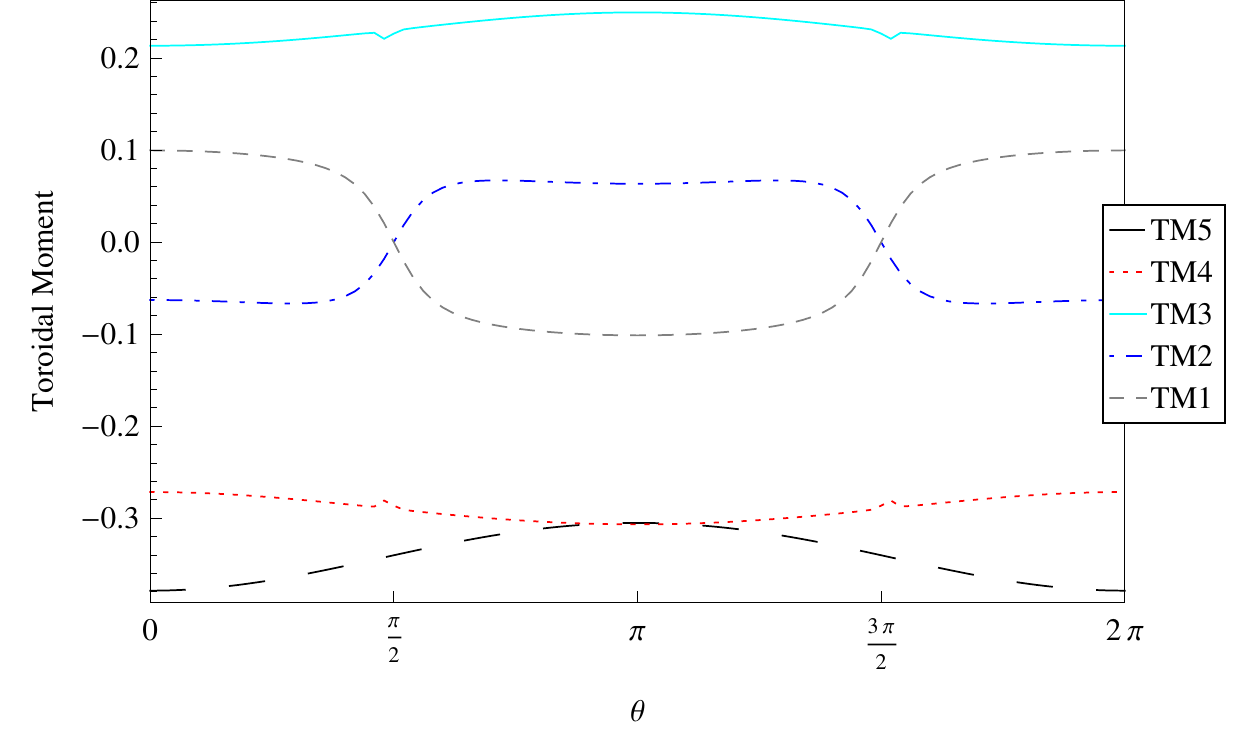}
\label{fig:subfig5a}}
\qquad
\subfloat[Subfigure 5b list of figures text][8-turn, circular toroidal helix with $R=1$, $a=0.5$, and $b=0.5$.]{
\includegraphics[width=0.4\textwidth]{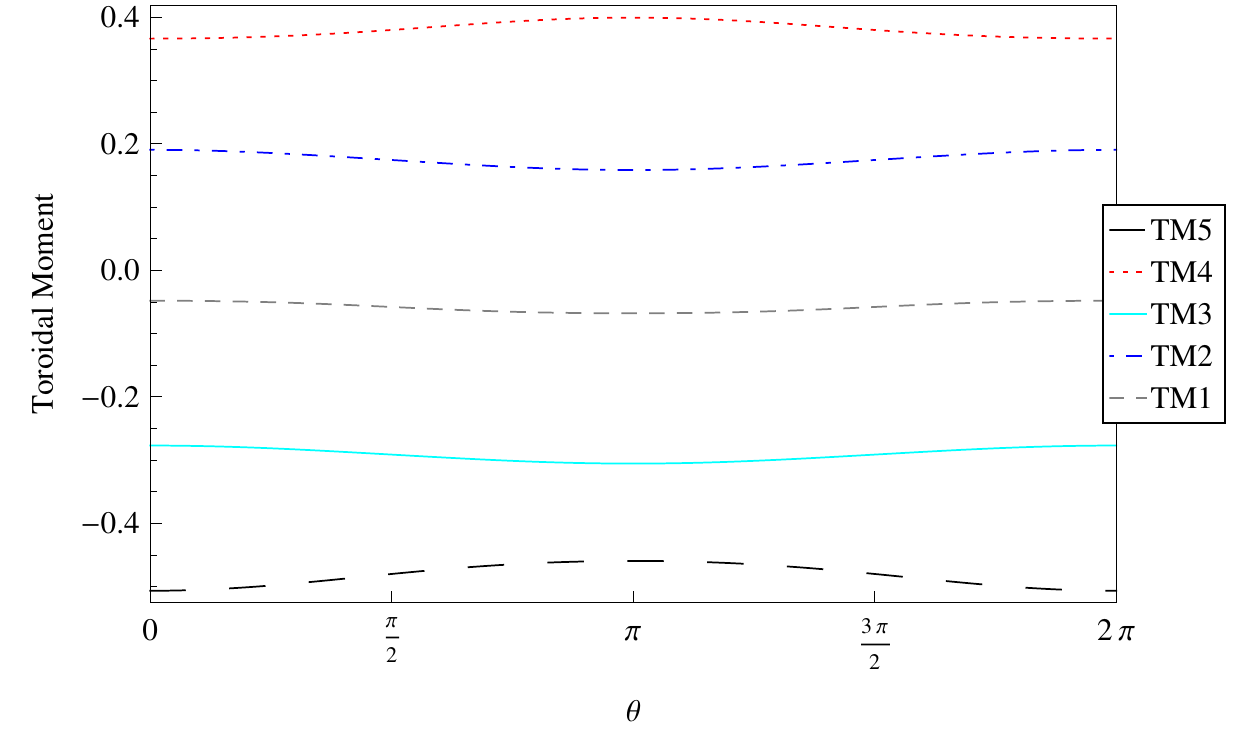}
\label{fig:subfig5b}}
\qquad
\caption{Toroidal moments (in units of $e \hbar R/m_e $) as a function of $\theta$ for the $p=2$ circular configurations.}
\label{fig:globfig5}
\end{figure}

\begin{figure}[h]
\centering
\subfloat[Subfigure 6a list of figures text][4-turn, tall elliptic toroidal helix with $a=0.25$, and $b=0.75$.]{
\includegraphics[width=0.4\textwidth]{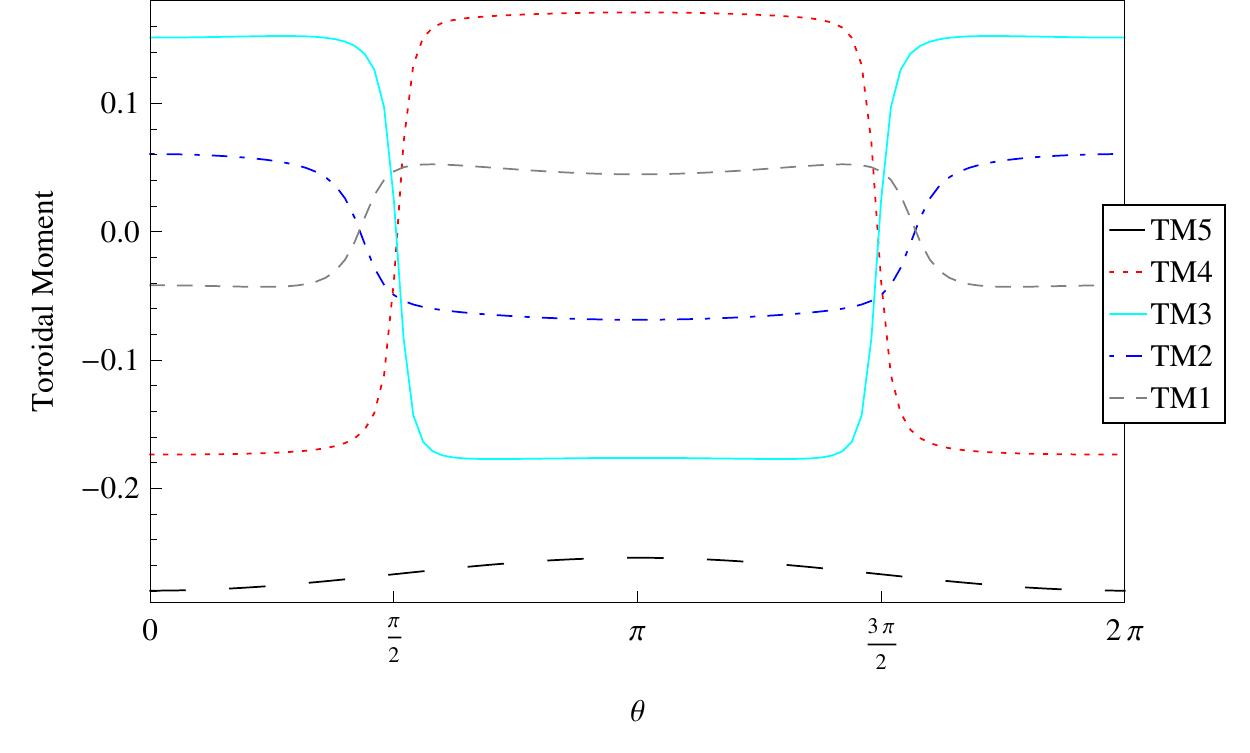}
\label{fig:subfig6a}}
\qquad
\subfloat[Subfigure 6b list of figures text][4-turn, flat elliptic toroidal helix with $a=0.75$, and $b=0.25$.]{
\includegraphics[width=0.4\textwidth]{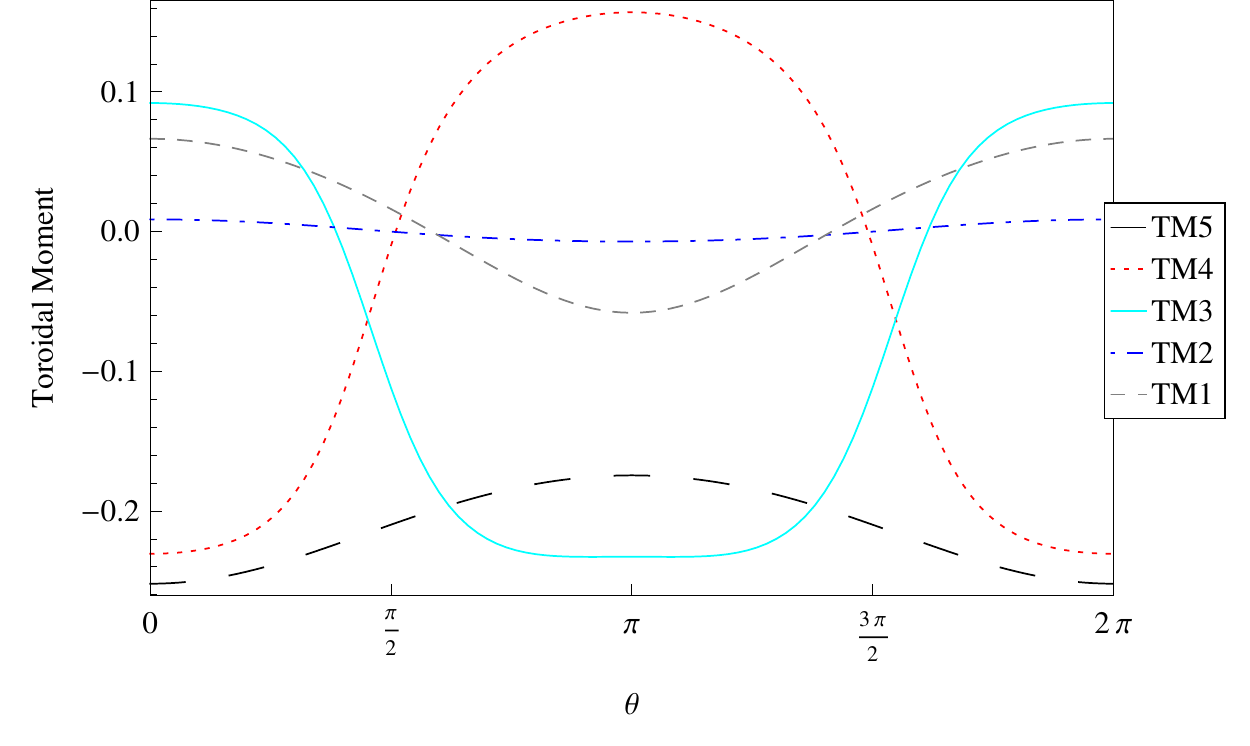}
\label{fig:subfig6b}}
\qquad
\subfloat[Subfigure 6c list of figures text][8-turn, tall elliptic toroidal helix with $a=0.25$, and $b=0.75$.]{
\includegraphics[width=0.4\textwidth]{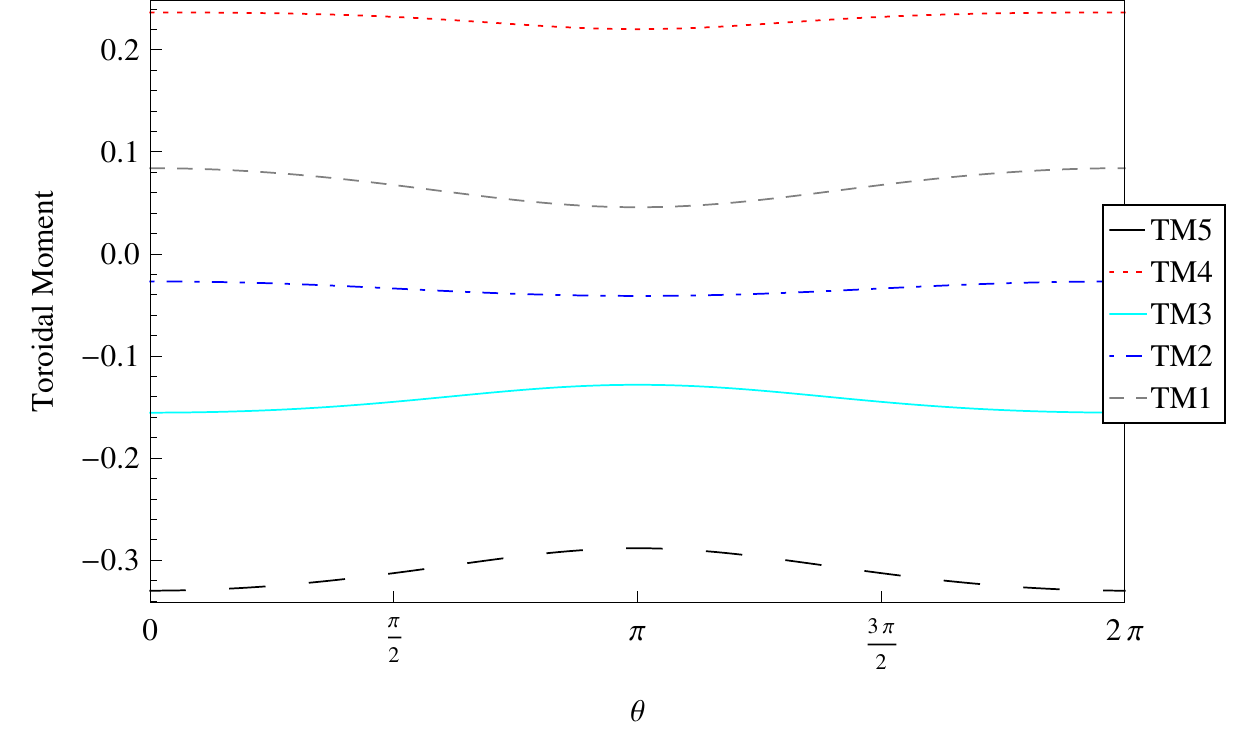}
\label{fig:subfig6c}}
\qquad
\subfloat[Subfigure 6d list of figures text][8-turn, flat elliptic toroidal helix with $a=0.75$, and $b=0.25$.]{
\includegraphics[width=0.4\textwidth]{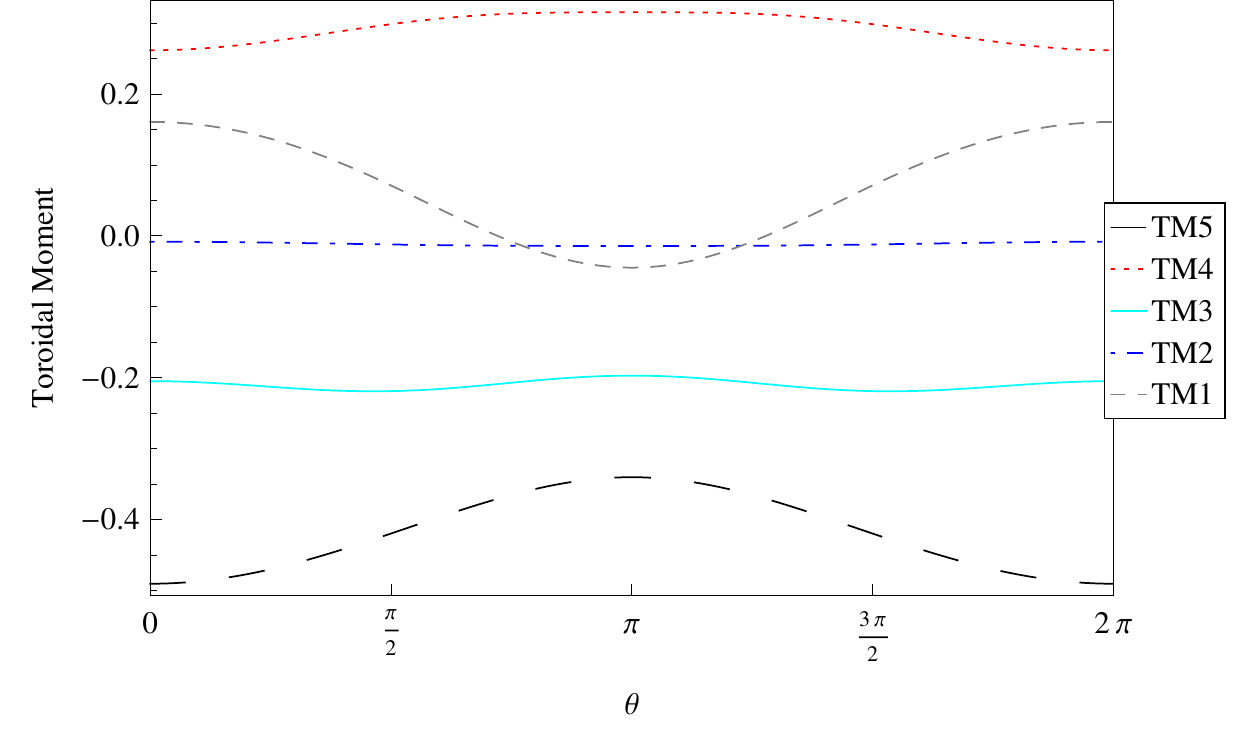}
\label{fig:subfig6d}}
\caption{Toroidal moments (in units of $e \hbar R/m_e $) as a function of $\theta$ for the $p=2$ elliptic configurations.}
\label{fig:globfig6}
\end{figure}

\begin{figure}[h]
\centering
\subfloat[Subfigure 7a list of figures text][4-turn, circular toroidal helix with $R=1$, $a=0.5$, and $b=0.5$.]{
\includegraphics[width=0.4\textwidth]{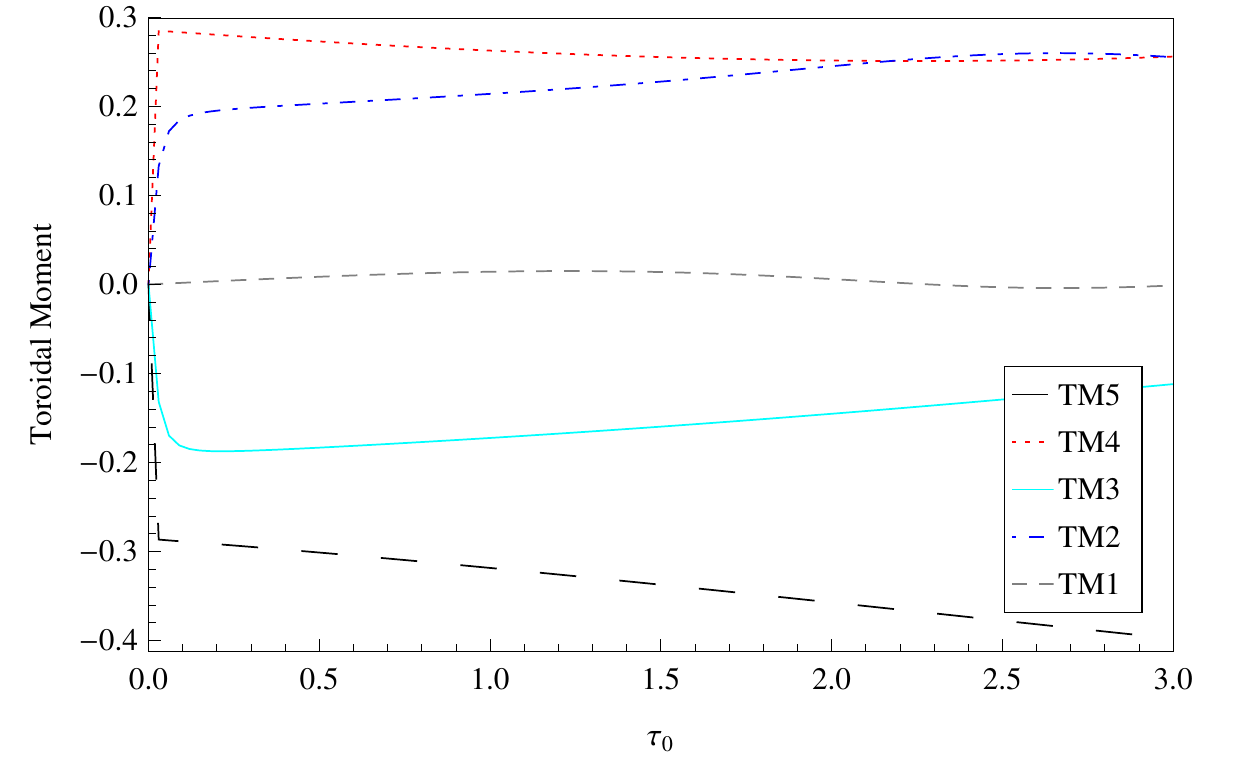}
\label{fig:subfig7a}}
\qquad
\subfloat[Subfigure 7b list of figures text][8-turn, circular toroidal helix with $R=1$, $a=0.5$, and $b=0.5$.]{
\includegraphics[width=0.4\textwidth]{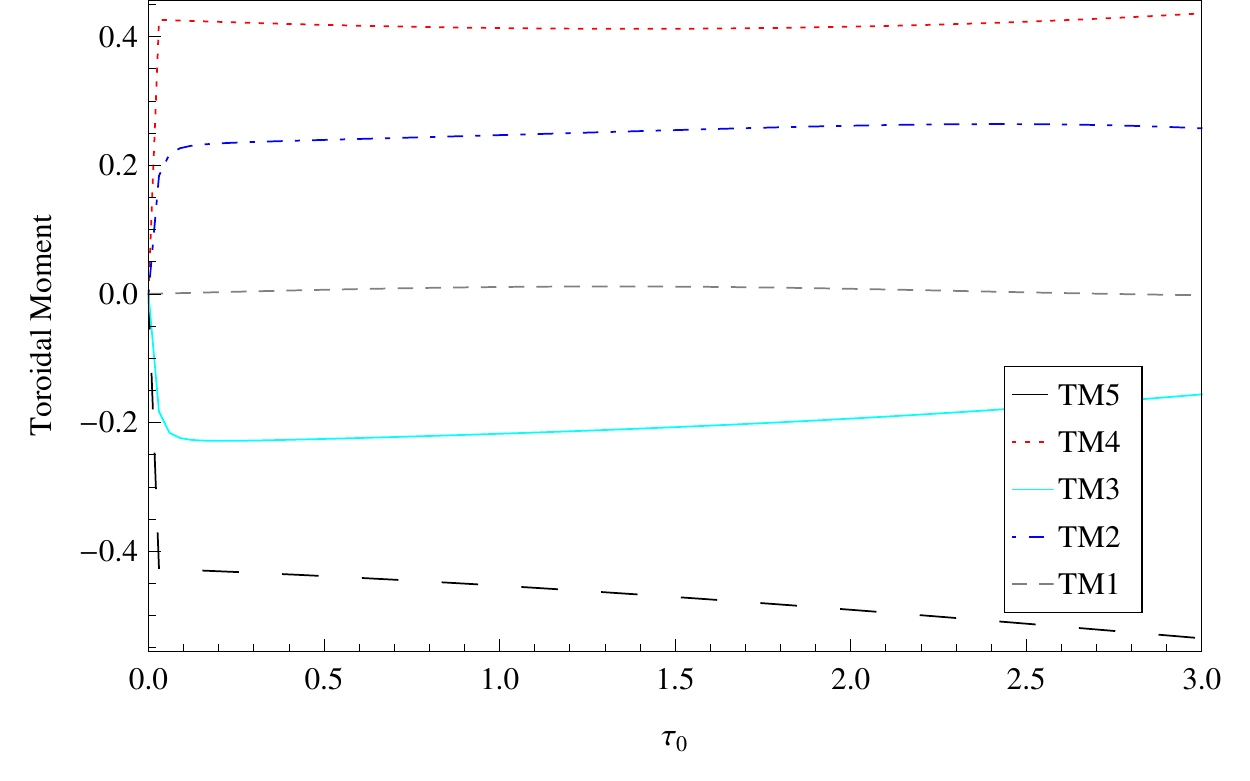}
\label{fig:subfig7b}}
\qquad
\caption{Toroidal moments (in units of $e \hbar R/m_e $) as a functin $\tau_0$ for the $p=2$ circular configurations.}
\label{fig:globfig7}
\end{figure}

\begin{figure}[h]
\centering
\subfloat[Subfigure 8a list of figures text][4-turn, tall elliptic toroidal helix with $a=0.25$, and $b=0.75$.]{
\includegraphics[width=0.4\textwidth]{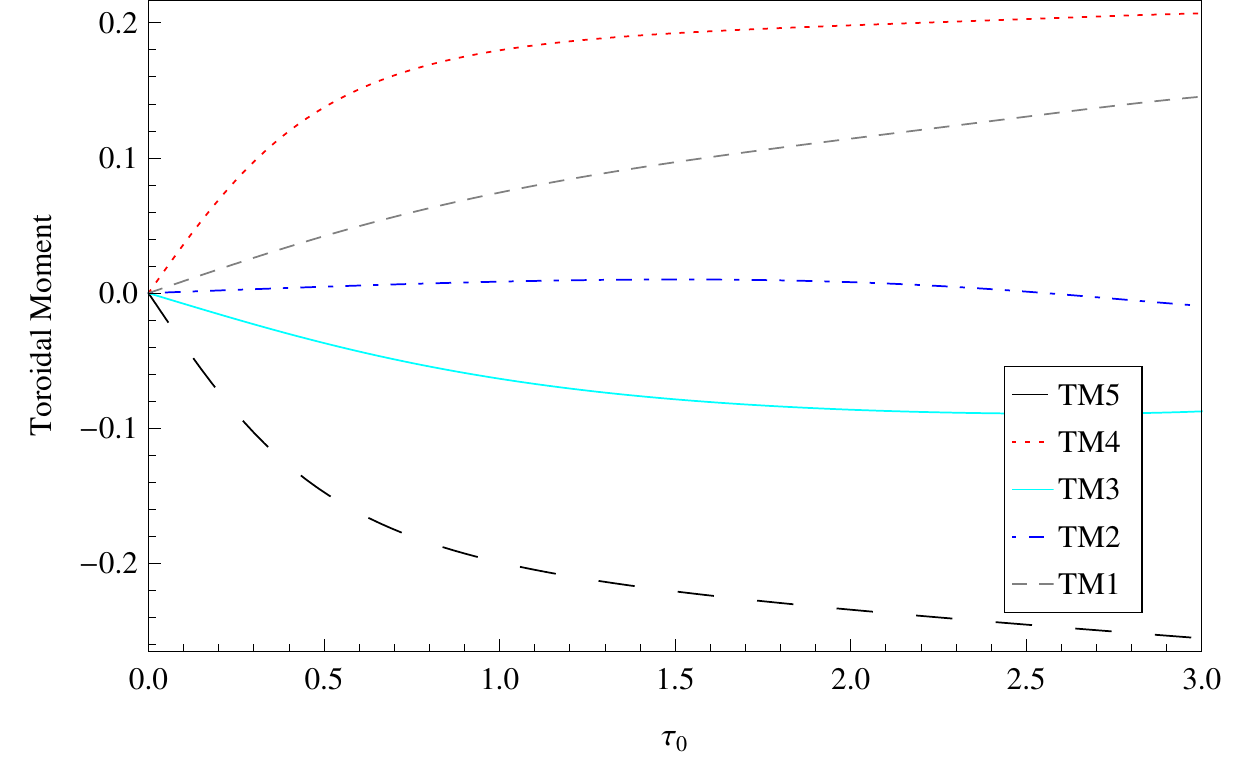}
\label{fig:subfig8a}}
\qquad
\subfloat[Subfigure 8b list of figures text][4-turn, flat elliptic toroidal helix with $a=0.75$, and $b=0.25$.]{
\includegraphics[width=0.4\textwidth]{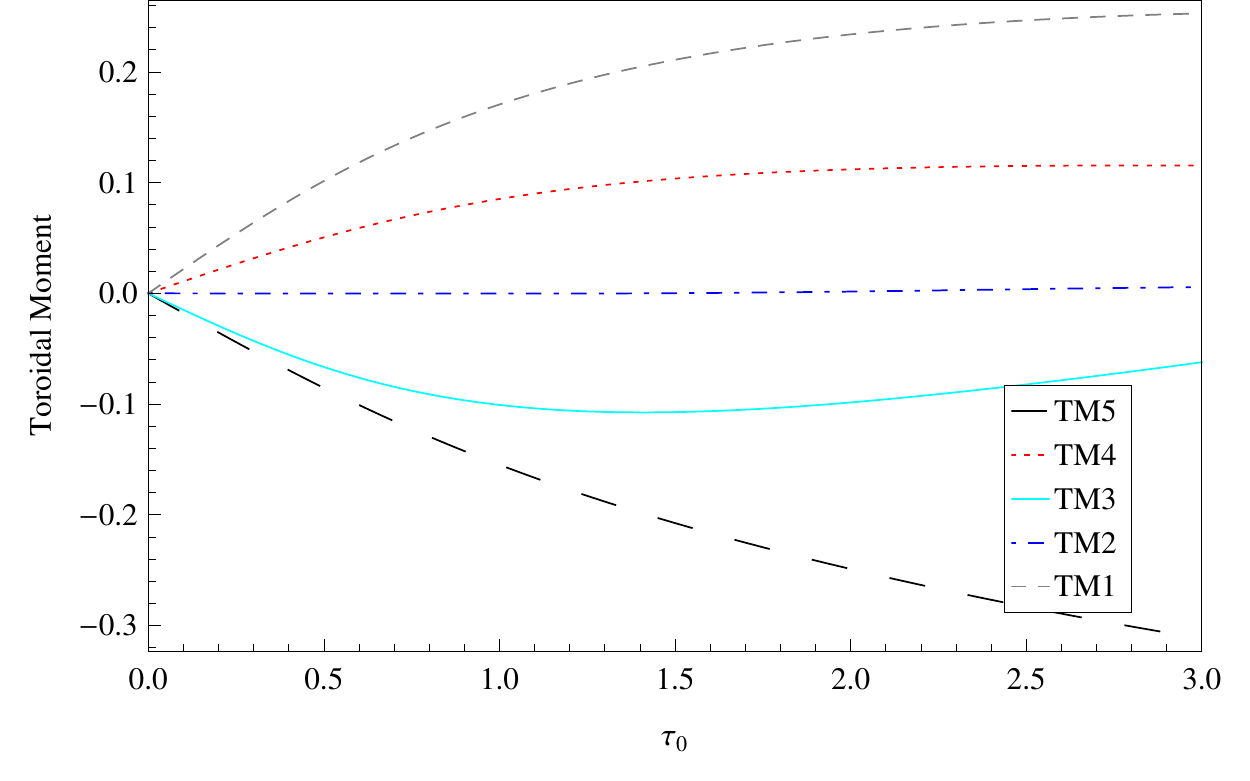}
\label{fig:subfig8b}}
\qquad
\subfloat[Subfigure 8c list of figures text][8-turn, tall elliptic toroidal helix with $a=0.25$, and $b=0.75$.]{
\includegraphics[width=0.4\textwidth]{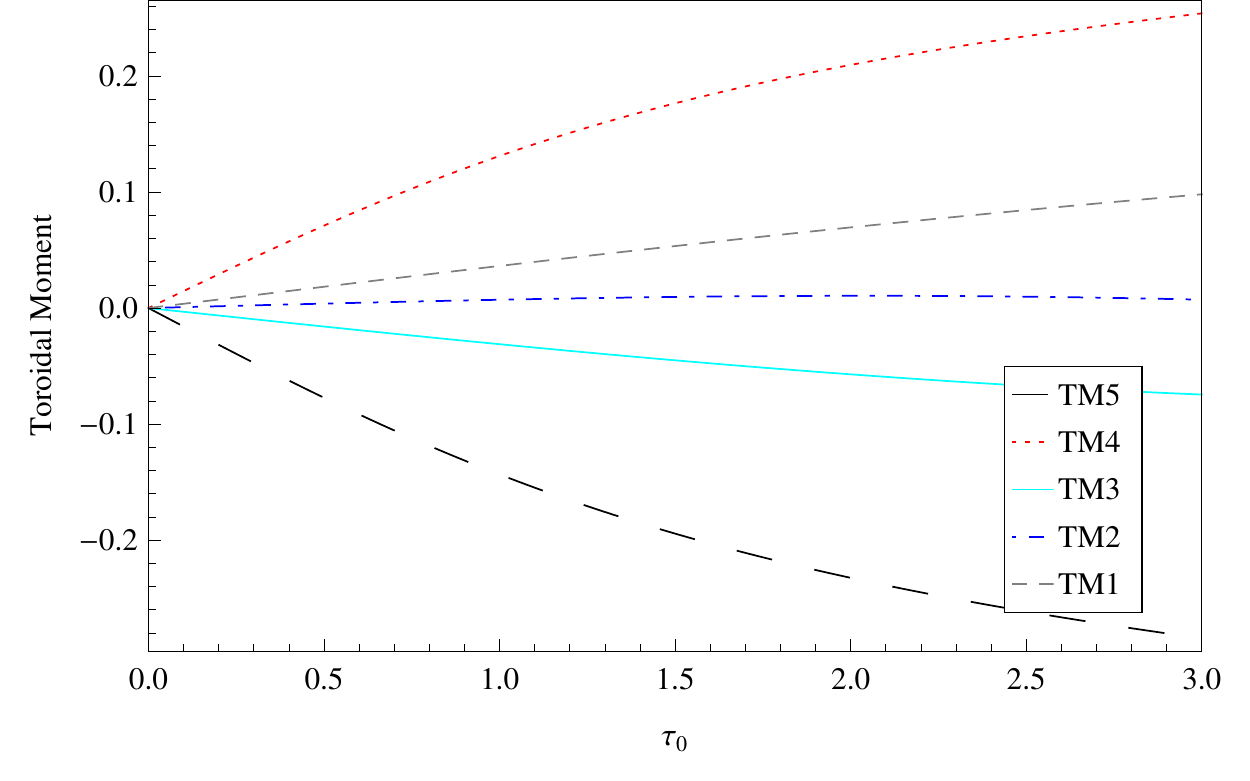}
\label{fig:subfig8c}}
\qquad
\subfloat[Subfigure 8d list of figures text][8-turn, flat elliptic toroidal helix with $a=0.75$, and $b=0.25$.]{
\includegraphics[width=0.4\textwidth]{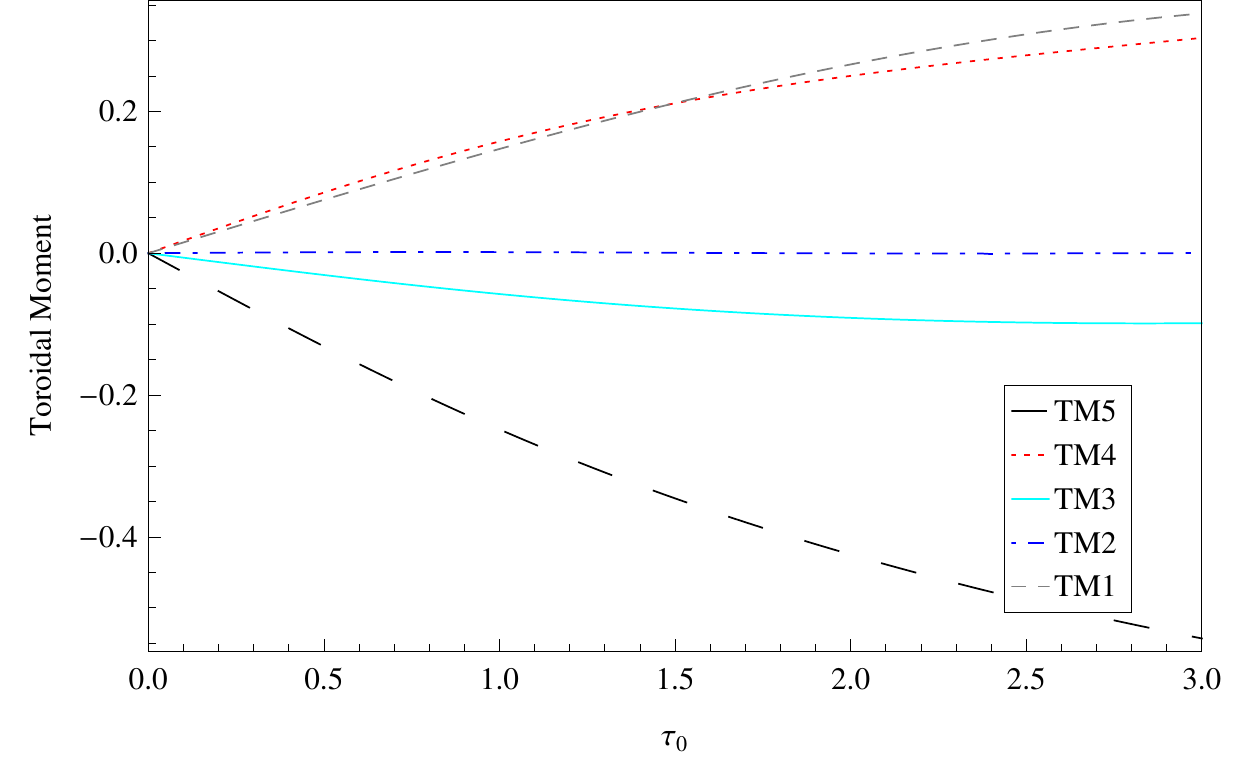}
\label{fig:subfig8d}}
\caption{Toroidal moments (in units of $e \hbar R/m_e $) for the $p=0$ elliptic configurations as a function of $\tau_0$.}
\label{fig:globfig8}
\end{figure}

\begin{figure}[h]
\centering
\subfloat[Subfigure 9a list of figures text][4-turn toroidal helix where $R=1$, $a=0.25$, and $b$ is varied from 0.1 to 0.9. The magnetic flux in the $z$-direction was held constant at $\tau_0=2$.]{
\includegraphics[width=0.4\textwidth]{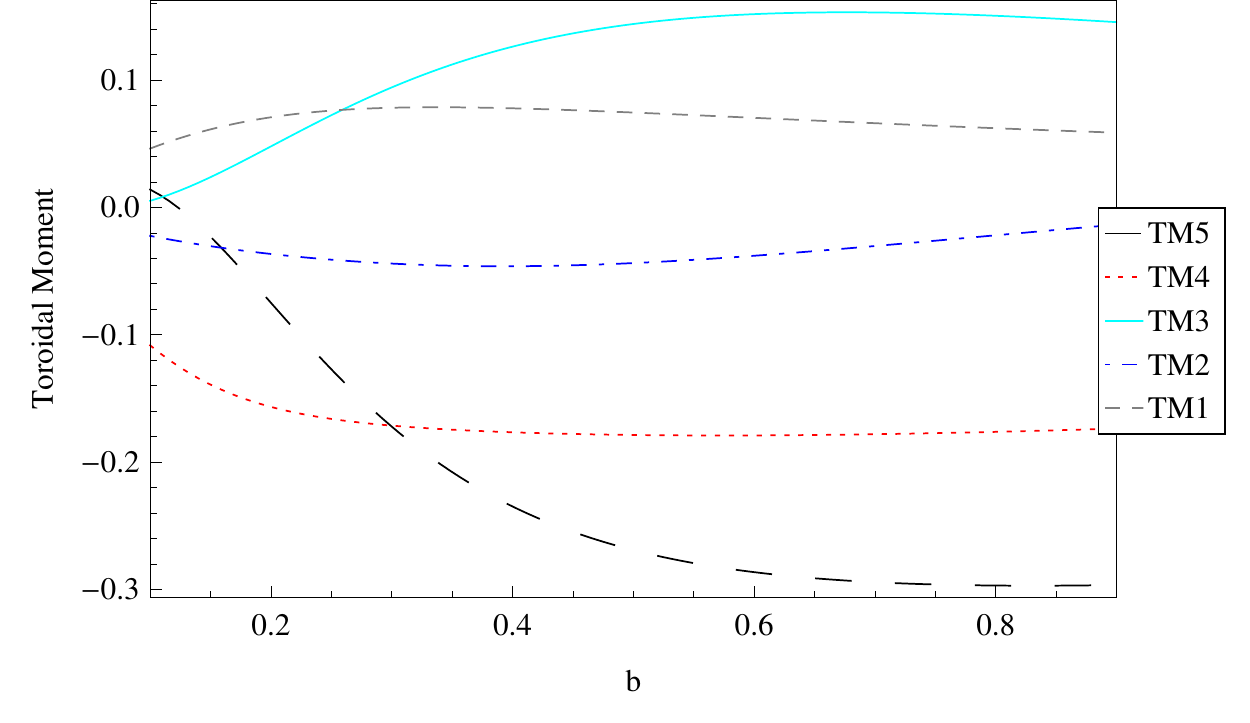}
\label{fig:subfig9a}}
\qquad
\subfloat[Subfigure 9b list of figures text][8-turn toroidal helix where $R=1$, $a=0.25$, and $b$ is varied from 0.1 to 0.9. The magnetic flux in the $z$-direction was held constant at $\tau_0=2$.]{
\includegraphics[width=0.4\textwidth]{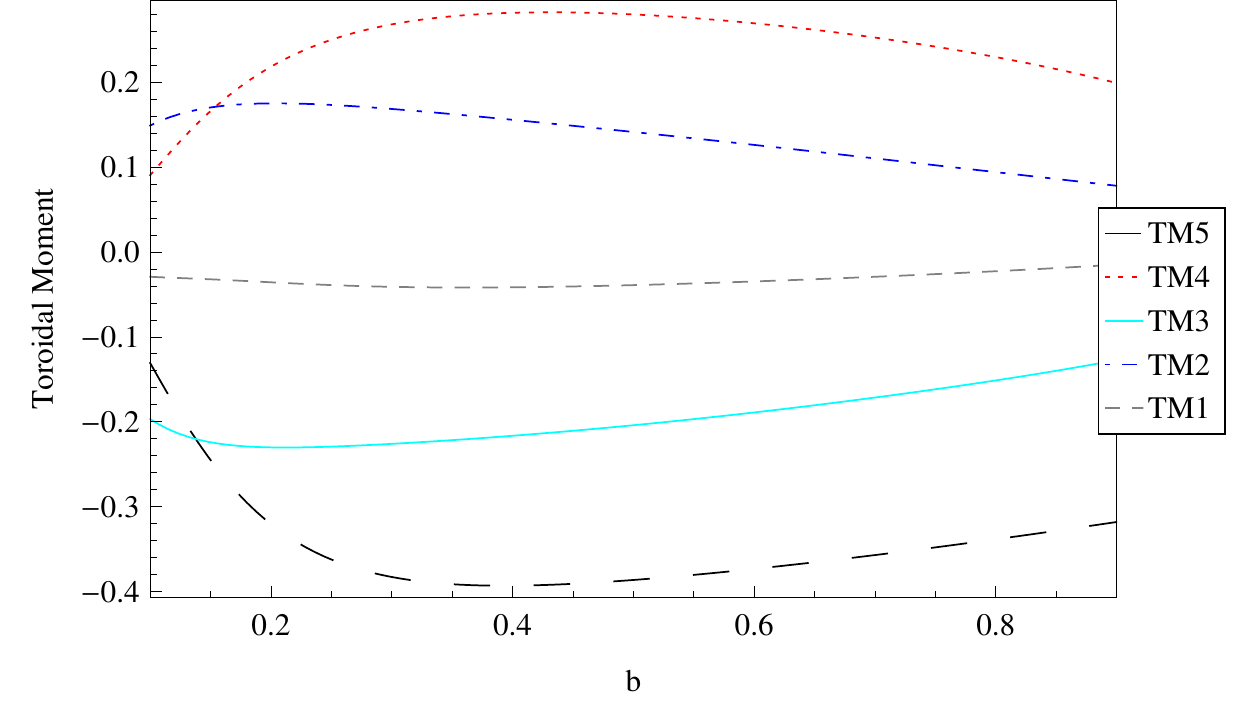}
\label{fig:subfig9b}}
\qquad
\caption{Toroidal moments (in units of $e \hbar R/m_e $) as a function of minor vertical axis, $b$, for the $p=2$ substate, with $\omega=4$ and $\omega=8$, and $a$ held constant.}
\label{fig:globfig9}
\end{figure}

\begin{figure}[h]
\centering
\subfloat[Subfigure 10a list of figures text][4-turn toroidal helix where $R=1$, $b=0.25$, and $a$ is varied from 0.1 to 0.9. The magnetic flux in the $z$-direction was held constant at $\tau_0=2$.]{
\includegraphics[width=0.4\textwidth]{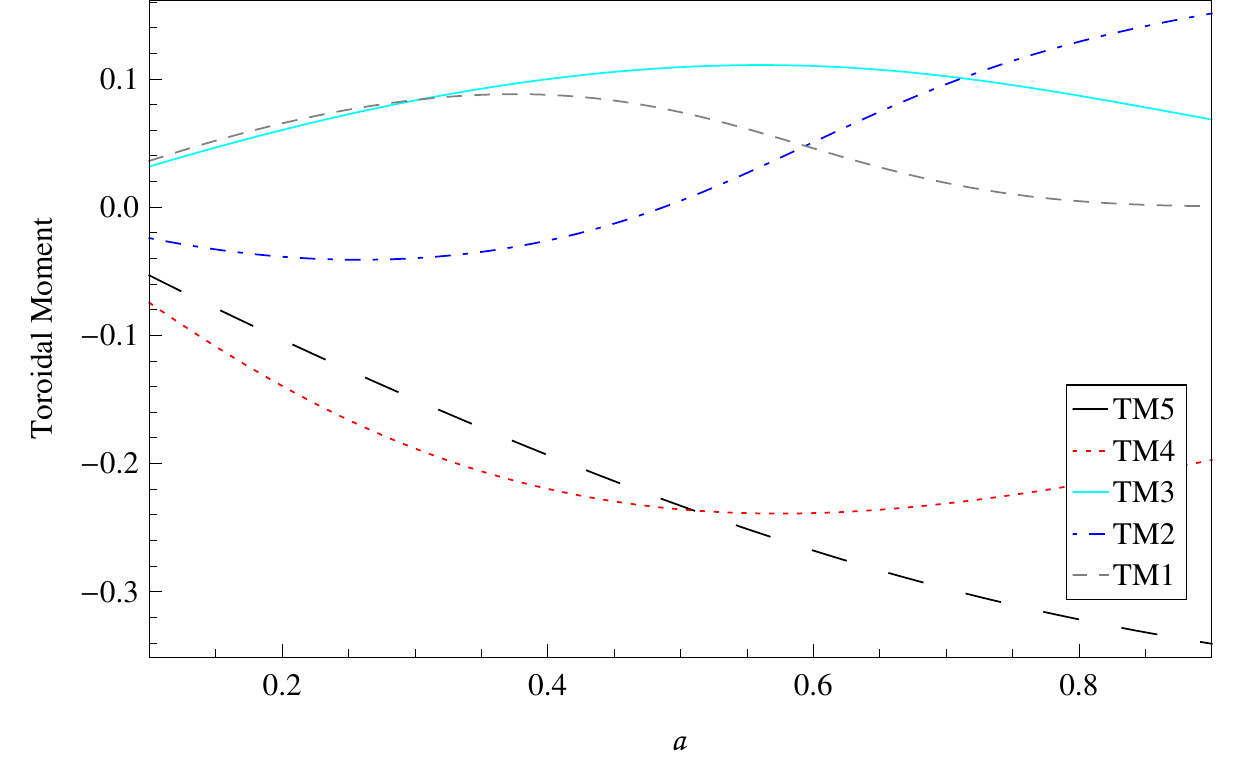}
\label{fig:subfig10a}}
\qquad
\subfloat[Subfigure 10b list of figures text][8-turn toroidal helix where $R=1$, $b=0.25$, and $a$ is varied from 0.1 to 0.9. The magnetic flux in the $z$-direction was held constant at $\tau_0=2$.]{
\includegraphics[width=0.4\textwidth]{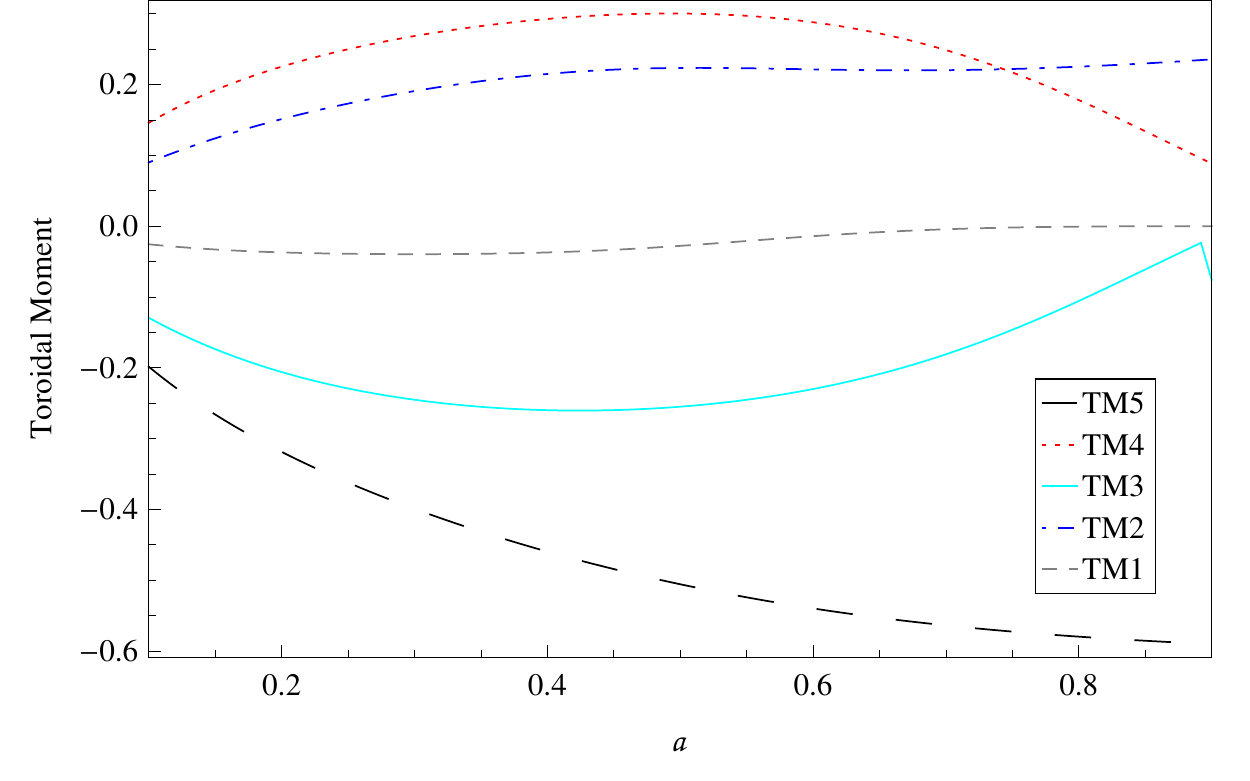}
\label{fig:subfig10b}}
\qquad
\caption{Toroidal moments (in units of $e \hbar R/m_e $) as a function of minor horizontal axis, $a$, for the $p=2$ substate, with $\omega=4$ and $\omega=8$ and constant $b$.}
\label{fig:globfig10}
\end{figure}

\end{document}